\documentclass[aps,prb,longbibliography,10pt
]{revtex4-2} 

\usepackage{graphicx,graphics,amsfonts,color,bm,amsmath,hyperref, 
amssymb,wasysym,ulem}
\usepackage[ansinew]{inputenc}

\def\be{\begin{equation}} 
\def\ee{\end{equation}} 
\def\ba{\begin{eqnarray}} 
\def\ea{\end{eqnarray}} 
\def\bc{\begin{center}} 
\def\ec{\end{center}}

\def\p{\partial}

\DeclareMathOperator\sign{sign}

\begin{document} 

\title{Theory of the in-plane photoelectric effect in two-dimensional electron systems} 

\author{S. A. Mikhailov} 
\email[Electronic mail: ]{sergey.mikhailov@physik.uni-augsburg.de} 
\affiliation{Institute of Physics, University of Augsburg, D-86135 Augsburg, Germany} 
\author{W. Michailow} 
\affiliation{Cavendish Laboratory, University of Cambridge, J.J.Thomson Avenue, Cambridge CB3 0HE, United Kingdom} 
\author{H. E. Beere} 
\affiliation{Cavendish Laboratory, University of Cambridge, J.J.Thomson Avenue, Cambridge CB3 0HE, United Kingdom} 
\author{D. A. Ritchie} 
\affiliation{Cavendish Laboratory, University of Cambridge, J.J.Thomson Avenue, Cambridge CB3 0HE, United Kingdom} 

\date{\today} 

\begin{abstract}
A new photoelectric phenomenon, the in-plane photoelectric (IPPE) effect, has been recently discovered at terahertz (THz) frequencies in a GaAs/Al$_x$Ga$_{1-x}$As heterostructure with a two-dimensional (2D) electron gas (W. Michailow et al., Sci. Adv. \textbf{8}, eabi8398 (2022)). In contrast to the conventional PE phenomena, the IPPE effect is observed at normal incidence of radiation, the height of the in-plane potential step, which electrons overcome after absorption of a THz photon, is electrically tunable by gate voltages, and the effect is maximal at a negative electron ``work function'', when the Fermi energy lies above the potential barrier. Based on the discovered phenomenon, efficient detection of THz radiation has been demonstrated. In this work we present a detailed theory of the IPPE effect providing analytical results for the THz wave generated photocurrent, the quantum efficiency, and the internal responsivity of the detector, in dependence on the frequency, the gate voltages, and the geometrical parameters of the detector. The calculations are performed for macroscopically wide samples at zero temperature. Results of the theory are applicable to any semiconductor systems with 2D electron gases, including III-V structures, silicon-based field effect transistors, and the novel 2D layered, graphene-related materials.
\end{abstract} 


\maketitle 

\tableofcontents

\section{Introduction\label{sec:intro}}

\begin{figure}[ht]
\includegraphics[width=8.5cm]{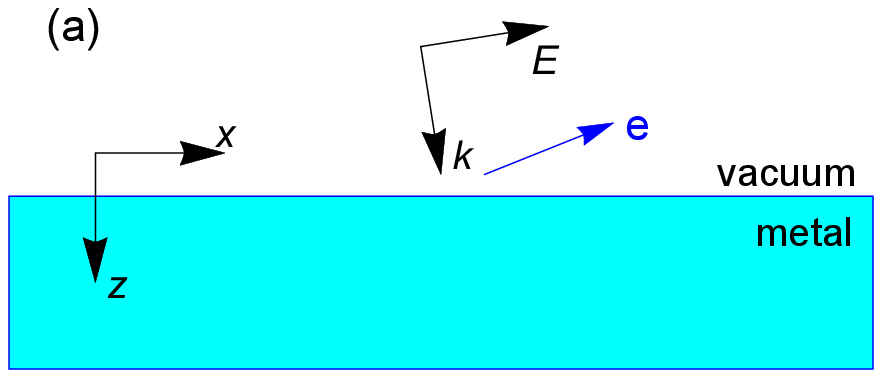} \\
\vspace{2mm}
\includegraphics[width=8.5cm]{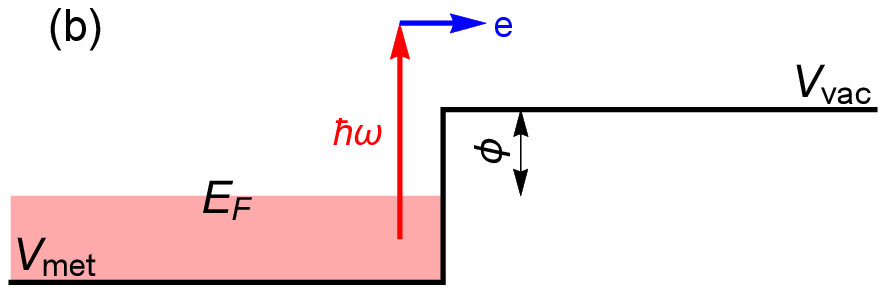}  \\
\vspace{2mm}
\includegraphics[width=8.5cm]{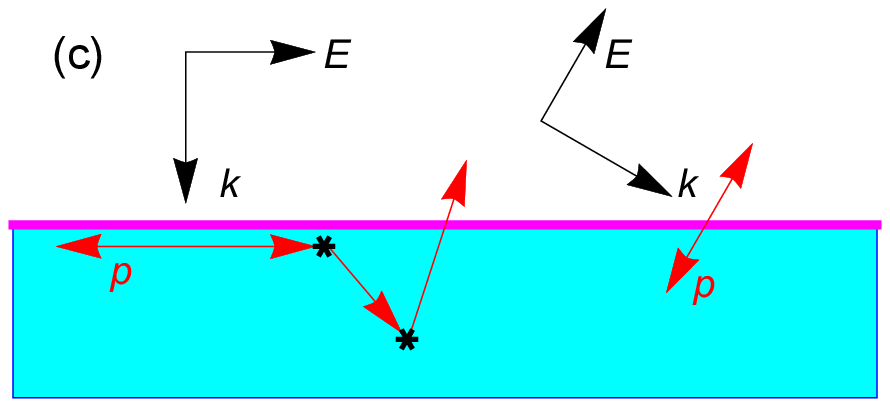}
\caption{\label{fig:physics-conventional} The conventional (external) photoelectric effect in a conducting medium. (a) The geometry of a typical experiment. (b) The band structure and the photon absorption process: $V_{\rm met}$ and $V_{\rm vac}$ are the bottoms of the conduction bands in a metal and in vacuum; $E_F$ is the Fermi energy of electrons in the metal. $\phi=V_{\rm vac}-E_F>0$ is the work function. (c) The dynamics of the photoexcitation process: At normal incidence of radiation electrons acquire a momentum $p_x$ \textit{parallel} to the surface, while to escape from the material they need a momentum component $p_z$ \textit{perpendicular} to the interface material-vacuum (shown by the thick magenta line). Electrons can get the momentum $p_z$ after a few scattering events in the metal or under oblique incidence of radiation.}
\end{figure}

In the conventional photoelectric effect, an electromagnetic wave irradiates a conducting medium, Figure \ref{fig:physics-conventional}(a), electrons absorb the light quanta, Figure \ref{fig:physics-conventional}(b), and acquire sufficient energy to overcome the built-in surface potential barrier $\phi$ and to escape from the material. The energy of the light quanta $\hbar\omega$ should exceed a certain value \cite{Lenard1902}, the material's work function $\phi$, defined as the difference between the lowest energy of an electron in vacuum $V_{\rm vac}$ and the Fermi level in the medium $E_F$, Figure \ref{fig:physics-conventional}(b). The maximum energy of the emitted photoelectrons equals  $E=\hbar\omega-\phi$, \cite{Einstein1905}. This process, the external photoelectric effect, takes place in the UV--Xray region of the electromagnetic spectrum, since the work functions of most metals lie in the range of several electronvolt. It can be used for generation of electricity from light, as well as for detection of electromagnetic radiation.

At lower frequencies, in the visible to near infrared ranges, a similar process results in the photovoltaic effect that takes place within an inhomogeneous semiconductor material. The incident radiation generates electron-hole pairs in the vicinity of a $p$-$n$-junction created within a semiconductor by different types of doping of the host material. The photoexcited electrons and holes are dragged in opposite directions due to the built-in electric field at the $p$-$n$-junction, thus generating the photovoltaic response. Here, the interband electronic transitions inside the semiconductor are used to generate the electrical photoresponse.

Moving toward even lower frequencies, mid- to far-infrared, the photon energy becomes smaller than the band gaps of semiconductors. Therefore, intraband transitions are used instead of interband photoexcitation. For example, in the hetero- or homojunction interfacial workfunction internal photoemission (HEIWIP/HIWIP) detectors \cite{Perera1992,Perera1995,Matsik2003,Perera2008,Shen2000,Lao2014} a potential step for electrons in the conduction band, similar to the one shown in Fig. \ref{fig:physics-conventional}(b), is created by different material content $(x)$ or different doping in 3D semiconductor heterostructures like GaAs/Al$_x$Ga$_{1-x}$As. These detectors work well at frequencies $f\sim 5-30$ THz (wavelengths $\sim 60-10$ $\mu$m) \cite{Perera2008}, but their detection efficiency keeps decreasing toward lower frequencies. One of the reasons is that the height of the potential step cannot be reduced down to a few meV since the practical values of the doping concentration and/or of the Al concentration $x$ cannot be made arbitrarily small \cite{Perera2008}; for example, a potential step of 4.1 meV (corresponding to 1 THz) would require the Al content ratio to be impractically small, $x\approx 0.003$. This limits the use of such detectors at frequencies on the order of several THz. In practice, the responsivity of HEIWIP/HIWIP detectors falls down by orders of magnitude when the radiation frequency approaches  frequencies $\sim 3$ THz from above \cite{Matsik2003,Perera2008,Bai2018}. 

Another crucial problem of detectors utilizing intraband electronic transitions is related to the transverse nature of electromagnetic waves. It can be explained by the example of the conventional (external) photoelectric effect in a metal, Fig. \ref{fig:physics-conventional}(c). The natural way of detecting electromagnetic radiation would be to send the light normally onto the material surface. However, under normal incidence of radiation, the electric field of the wave $\bm E$, and hence the momentum $\bm p$ that electrons obtain from the wave, are parallel to the surface. Electrons may obtain a very large energy from the incident electromagnetic wave, but this does not help them to overcome the potential barrier since they move parallel to the interface, \cite{Tamm1931,Mitchell1934}. This results in a low photoelectron yield. To mitigate this problem one can either use oblique incidence of $p$-polarized waves or rely on scattering processes which could enable electrons to get momentum components normal to the interface, Fig. \ref{fig:physics-conventional}(c). Both cases are not optimal; in addition, scattering is a non-deterministic, random process that reduces the energy of photoexcited electrons and thus diminishes the efficiency. Moreover, it limits the intrinsic response time of the effect to scattering times. The described problem is also present in HEIWIP and HIWIP detectors, as well as in other types of infrared photodetectors based on low-dimensional electron systems in semiconductors, for example, in quantum well \cite{Levine1993}, quantum dot \cite{Stiff2009}, and TACIT (tunable antenna-coupled intersubband terahertz, Ref. \cite{Cates1998}) photodetectors. 

\begin{figure}[ht]
\includegraphics[width=8.5cm]{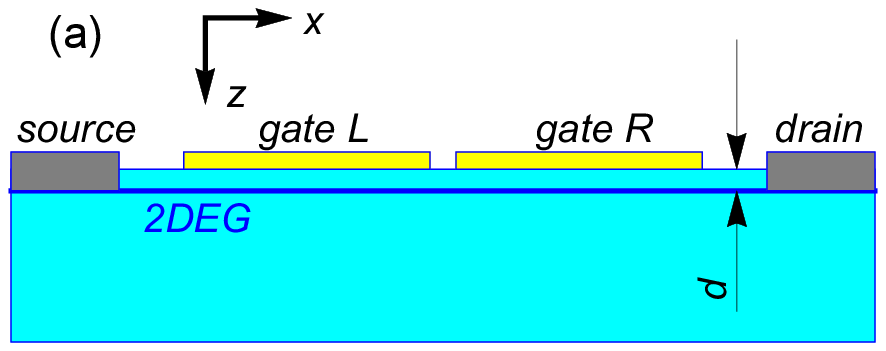}  \\
\vspace{2mm}
\includegraphics[width=8.5cm]{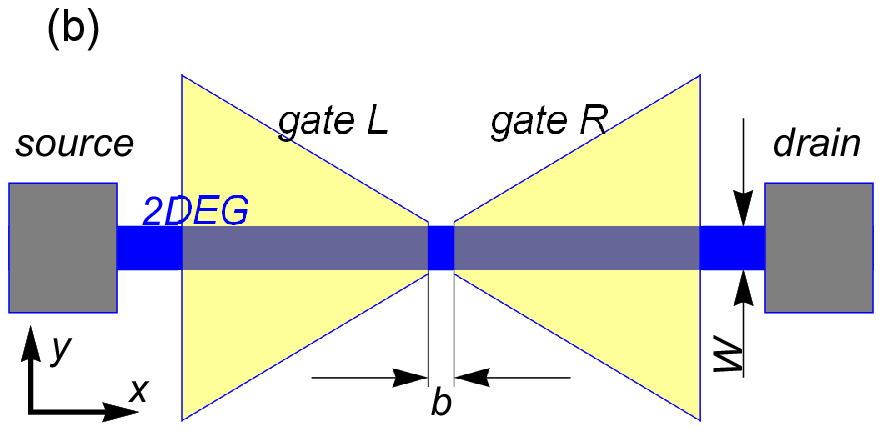}  \\
\vspace{2mm}
\includegraphics[width=8.5cm]{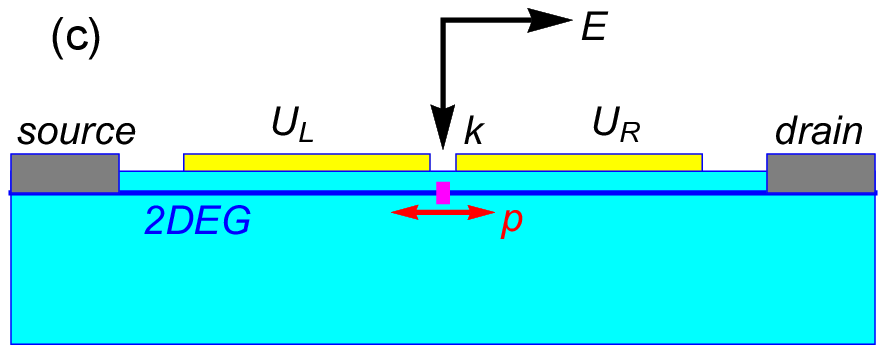} \\
\vspace{2mm}
\includegraphics[width=8.5cm]{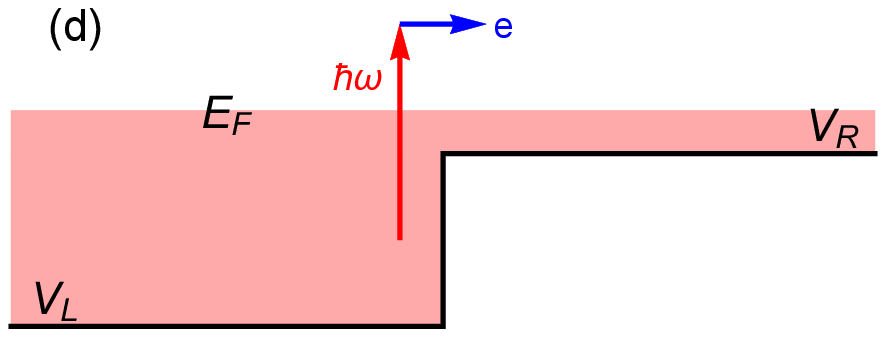} 
\caption{\label{fig:physics} The geometry and the operation principle of the detector based on the in-plane photoelectric effect. (a) Side and (b) top view of the detector: a 2DEG channel of the width $W$ is connected to source and drain contacts and covered by two (left and right) gates. The gates have the shape of an antenna separated by a gap of the width $b$. The 2D gas lies at the depth $d$ under the surface of the sample. (c) The operation mode of the detector: different voltages $U_{L}$ and $U_{R}$ are applied to the left and right gates thus creating a potential step in the 2DEG plane on the line $x=0$ (illustrated by the magenta rectangle) in the lateral ($x$-) direction, on the way of electrons from one contact to another. Irradiation of the structure by the normally incident THz waves leads to a strong ac electric field focused under the gap between the gates; electrons in this area obtain a large oscillating momentum $\bm p$ (illustrated by red arrows) parallel to the 2DEG plane and perpendicular to the potential step. (d) The potential step $V_L\Theta(-x)+V_R\Theta(x)$ created by different voltages $U_{L}$ and $U_{R}$ on the gates and the photon absorption process in the vicinity of the potential step. The Fermi level $E_F$ may lie above both conduction band bottoms $V_L$ and $V_R$. The maximum photocurrent was observed in Ref. \cite{Michailow22} when the quantity $\phi=V_R-E_F$ -- the analogue of the work function in the conventional photoelectric effect -- was negative, $E_F>V_R$.}
\end{figure}

Recently, a new type of photoelectric effect has been discovered in our paper, Ref. \cite{Michailow22}. The effect was observed in a GaAs-Al$_x$Ga$_{1-x}$As heterostructure with a two-dimensional (2D) electron gas (EG) lying under the semiconductor surface at a distance $d$, Fig. \ref{fig:physics}(a). The 2D channel of width $W$ in the $y$-direction, had source and drain contacts and was covered by two, left and right, gates, Fig. \ref{fig:physics}(b). The gates had the shape of a bow-tie antenna and focused the incident THz radiation into the gap of the width $b$ between the gates (the shape of the gates was more complicated in Ref. \cite{Michailow22}, but we will omit unnecessary details and consider a simpler model). The width $b$ was substantially smaller than the mean free path of 2D electrons $l_{\mathrm{mfp}}$ in the channel, $b\ll l_{\mathrm{mfp}}$. The application of different dc voltages $U_{L}$ and $U_{R}$ to the left and right gates generated an electrically tunable potential step for electrons moving in the 2D channel between the source and the drain, Fig. \ref{fig:physics}(d). 

When the device was exposed to normally incident electromagnetic radiation (with the frequency $f=1.9$ THz in Ref. \cite{Michailow22}), the electric field of the wave $E_x$, focused in the narrow gap between the antenna wings, caused the electrons to oscillate with the momentum $p_x$ parallel to the surface, Fig. \ref{fig:physics}(c). This time, however, the potential step was created not in the vertical, $z$-direction, but in the lateral, $x$-direction. Electrons absorbed THz photons in the gap between the gates and jumped onto the step, Fig. \ref{fig:physics}(d), generating an electron flow in the 2DEG plane in the lateral direction, from the area of a large electron density to the area of a low electron density. The problem caused by the transverse nature of the electromagnetic waves was thus solved in Ref. \cite{Michailow22} in a simple and elegant way: the potential step, which lies in the plane $z=0$ in the conventional photoelectric effect (magenta line in Fig. \ref{fig:physics-conventional}(c)), now lies inside the 2D electron layer on the line $x=0$ (magenta rectangle in Fig. \ref{fig:physics}(c)). The ideal conditions for the generation of photocurrent in the 2D gas in the lateral direction have thus been created. This phenomenon was called the \textit{in-plane photoelectric} (IPPE) effect. As shown in Ref. \cite{Michailow22}, it enables highly sensitive detection of THz radiation, and the corresponding device was called a photoelectric tunable-step (PETS) detector. It was also shown there that other (classical) detection mechanisms, such as the bolometric \cite{DeglInnocenti17}, photothermoelectric \cite{Gabor11,Viti16,Castilla19,Viti20} effects, plasmonic mixing \cite{Dyakonov96,Viti15,Bandurin18,Sun12a}, etc., can not explain the experimental findings of Ref. \cite{Michailow22}: they are not applicable or give a smaller photocurrent than experimentally observed.

As was discussed in Ref. \cite{Michailow22}, there is another aspect that significantly distinguishes the IPPE phenomenon from the conventional photoelectric effect. The condition $E=\hbar\omega-\phi$ for the energy of the photoemitted electrons, introduced by Einstein \cite{Einstein1905}, seems to imply that the lower the radiation frequency, the smaller the work function $\phi$ should be. It is this condition that represented severe difficulties in realizing the HEIWIP and HIWIP detectors at THz frequencies, since it is difficult to create a potential step of a few meV height in 3D semiconductor homo- or heterojunctions. However, the experiment \cite{Michailow22} showed that the quantity $\phi$, the ``work function'', does not need to be positive: the effect was observed and, moreover, it was maximal, when the 2D electron gas was degenerate ($E_F>V_L$, $E_F>V_R$) on both sides of the potential step at $x=0$, Fig. \ref{fig:physics}(d). The opportunity to electrically tune the height of the potential step $V_R-V_L$ completes the list of great features of the IPPE effect. 

In this paper we present a detailed analytical theory of the IPPE effect at zero temperature $T=0$ and in macroscopically wide ($N_{1D}\gg 1$) samples; here $N_{1D}$ is the number of 1D subbands in the 2D channel. In Section \ref{sec:formulation} we introduce the main approximations of our model and formulate the time-dependent Schr\"odinger equation which has to be solved. In Section \ref{sec:0thorder} we solve this equation in the zeroth order of perturbation theory. Section \ref{sec:1storder} contains the main results of our work: we solve the photoresponse problem within the first-order perturbation theory and calculate different physical quantities characterizing the operation of the PETS detector, such as e.g. the quantum efficiency, internal quantum resistance, and responsivity. Finally, in Section \ref{sec:conclusion} we summarize our results. 

\section{Formulation of the problem\label{sec:formulation}}

Consider the structure shown in Figure \ref{fig:physics}(a,b). If no voltages are applied to the gates, the bottom of the conduction band and the equilibrium chemical potential of electrons $\mu_0=E_F$ do not depend on the coordinate $x$. If dc voltages $U_L$ and $U_R$ are applied to the left and right gates, respectively, the potential energy $V_0(x)=V_0(x;V_L,V_R)$ seen by 2D electrons in the channel acquires the form of a smooth step function which varies from $V_L$ at $x\to-\infty$ to $V_R$ at $x\to+\infty$ on a scale of the order of $\max\{b,d\}\ll l_{\mathrm{mfp}}$. Due to the screening of the external potential by electrons in the 2D gas, the heights of the potential energy seen by electrons, $V_L$ and $V_R$, are related to the gate voltages $U_L$ and $U_R$ by the formula
\be 
V_{L,R}=\frac{-eU_{L,R}}{\epsilon(q,\omega=0)}\label{2Dscreening},\ \ \ \epsilon(q,\omega=0)=1+\frac {4d}{a_B}
\ee
where $\epsilon(q,\omega=0)$ is the static dielectric function of the 2D electron gas \cite{Stern67,Chaplik72} and $a_B$ is the effective Bohr radius (in GaAs $a_B\approx 10$ nm). Equation (\ref{2Dscreening}) can be derived within the local capacitance approximation under the assumption that nowhere under the gates the electron gas is depleted.

In the experiment the metallic gates simultaneously serve as antenna wings and the structure is irradiated by THz waves. Since the THz frequency $\omega=2\pi f$ is much smaller than the plasma frequency in metals, the metallic gates can be considered as quasi-equipotential at the frequency $\omega$. As a result, the influence of THz radiation can be described by a periodic increase and decrease of the gate potentials $U_L$ and $U_R$, 
\be 
U_{L,R}\to U_{L,R} \pm \Delta U(t), \ \ \ \ \Delta U(t)=\frac 12\Delta\Phi_{ac}(t),
\ee
where $\Delta\Phi_{ac}(t)\propto \cos\omega t$ is the potential difference between the left and right antenna wings resulting from the THz irradiation. The asymptotic potential energies $V_L$ and $V_R$ also oscillate, $V_{L,R}\to V_{L,R} \pm \Delta V(t)$. Notice that at high (THz) frequencies the oscillation amplitudes $\Delta V(t)$ and $\Delta U(t)$ are related by the formula $\Delta V(t)=-e\Delta U(t)/\epsilon(q,\omega)\approx -e\Delta U(t)$, since the dynamic dielectric function of the 2D electron gas $\epsilon(q,\omega)$ is close to 1 at THz frequencies. The $x$-dependence of the ac potential $V_1(x,t)$ acting on the electrons can then be described by the same smooth function which determines $V_0(x)$ but with time-dependent asymptotes $V_L(t)$ and $V_R(t)$. The motion of electrons in the 2D channel is thus determined by the time-dependent Schr\"odinger equation
\be 
i\hbar \frac{\p \Psi}{\p t}=\hat H\Psi=\hat H_0\Psi+\hat H_1(x,t)\Psi
\label{SE}
\ee
where the Hamiltonian $\hat H$ consists of the unperturbed part 
\be 
\hat H_0=-\frac{\hbar^2}{2m}\frac{\p^2}{\p x^2}-\frac{\hbar^2}{2m}\frac{\p^2}{\p y^2}+V_0(x),
\ee
and of the perturbation $\hat H_1=V_1(x,t)$. We will also assume that in the $y$-direction the channel is confined by infinitely high potential walls at $y=0$ and $y=W$, so that the wavefunction satisfies the boundary conditions $\Psi|_{y=0}=\Psi|_{y=W}=0$.

To simplify the problem and to get analytical results for the photoresponse of the 2D channel we now replace the true, smooth potentials $V_0(x)$ and $V_1(x,t)$ by step-like functions, Fig. \ref{fig:physics}(d). Thus we assume that the potential energy of electrons in the 2D channel is
\be 
V_0(x)=V_L+(V_R-V_L)\Theta(x),
\label{V0}
\ee
where $\Theta(x)$ is the Heaviside function, and the additional ac potential energy due to the THz irradiation has the form 
\be 
V_1(x,t)=\frac 12e\Delta\Phi_{\rm ac}\sign(x)\cos\omega t.
\label{V1}
\ee
Here $\Delta\Phi_{\rm ac}$ is the amplitude of the ac potential difference between the antenna wings, which can be evaluated as $\Delta\Phi_{ac}\simeq E_{ac}b$, where $E_{ac}$ is the average of the ac electric field in the plane of the 2DEG under the gap between the two gates. We will assume, without loss of generality, that $V_R>V_L$. Now we solve the problem (\ref{SE}) by applying the perturbation theory in the zeroth and first orders in $V_1$.

\section{Zeroth-order approximation\label{sec:0thorder}}

\subsection{Wave functions and transmission coefficients\label{sec:0thorderTrans}}

In the zeroth order, the equation to be solved is
\be 
i\hbar \frac{\p \Psi^{(0)}(x,y,t)}{\p t}=\hat H_0\Psi^{(0)} =-\frac{\hbar^2}{2m}\frac{\p^2 \Psi^{(0)}(x,y,t)}{\p x^2}-\frac{\hbar^2}{2m}\frac{\p^2 \Psi^{(0)}(x,y,t)}{\p y^2}+V_0(x)\Psi^{(0)}(x,y,t)\label{0thOrder}.
\ee
Its solution can be characterized by two parameters, the total energy $E$ and the 1D subband index $n$ ($=1,2,\dots$), and has the form 
\be 
\Psi^{(0)}(x,y,t)=e^{-iEt/\hbar}\sin\frac{\pi n y}{W}\psi^{(0)}_{En}(x).
\ee
where the function $\psi^{(0)}_{En}(x)$ satisfies the standard one-dimensional Schr\"odinger equation
\be 
E \psi^{(0)}_{En}(x) =-\frac{\hbar^2}{2m}\frac{\p^2 \psi^{(0)}_{En}(x)}{\p x^2}+E_Wn^2\psi^{(0)}_{En}(x)+V_0(x)\psi^{(0)}_{En}(x),
\label{0thOrderpsi(x)}
\ee
with
\be 
E_W=\frac{\hbar^2\pi^2}{2mW^2}
\ee
being the transverse quantization energy due to the confinement potential in the $y$-direction.

\begin{figure}
\includegraphics[width=0.49\textwidth]{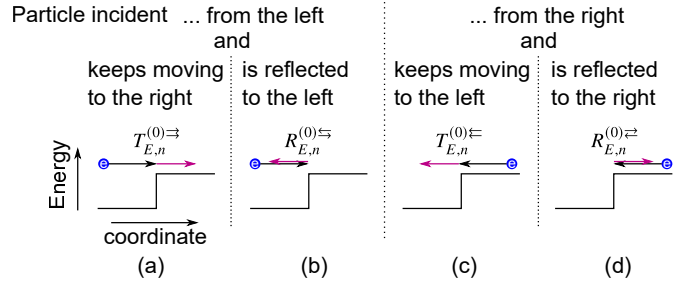}
\caption{\label{fig:DoubleArrows} The definition of the double-arrow notation in Eqs. (\ref{tr->})--(\ref{R0}): (a) electrons incident on the potential step from the left that keep moving to the right, (b) incident from the left and being reflected to the left, (c) incident from the right that keep moving to the left, and (d) incident from the right and being reflected to the right.}
\end{figure}

Consider first the energies $E$ above both potential barriers, $E-E_Wn^2>V_R>V_L$. Then for the particles running to the right the wave function is 
\be 
\psi^{(0)\Rightarrow}_{En}(x)=\left\{ 
\begin{array}{lr}
e^{iQ(E-E_Wn^2-V_L)x}+r^{(0)\leftrightarrows}_{En}e^{-iQ(E-E_Wn^2-V_L)x}, &\textrm{ if } x<0,\\
t^{(0)\rightrightarrows}_{En}e^{iQ(E-E_Wn^2-V_R)x}, & \textrm{ if } x>0,\\
\end{array}
\right.
\label{PsiRight}
\ee
and for the particles running to the left it is 
\be 
\psi^{(0)\Leftarrow}_{En}(x)=\left\{ 
\begin{array}{lr}
t^{(0)\leftleftarrows}_{En}e^{-iQ(E-E_Wn^2-V_L)x}, & \textrm{ if } x<0,\\
e^{-iQ(E-E_Wn^2-V_R)x}+r^{(0)\rightleftarrows}_{En}e^{iQ(E-E_Wn^2-V_R)x}, &\textrm{ if } x>0.\\
\end{array}
\right.
\label{PsiLeft}
\ee
The function $Q(E)$ here is the momentum of particles with the energy $E$, 
\be 
Q(E)=\frac{\sqrt{2mE}}\hbar;
\ee
this is a complex-valued function of energy defined so that if $E$ is negative, then \be 
Q(E)=+i\frac{\sqrt{2m(-E)}}\hbar.
\label{Qof-E}
\ee 
The transmission and reflection amplitudes $t^{(0)\rightrightarrows}_{En}$, $r^{(0)\leftrightarrows}_{En}$, $t^{(0)\leftleftarrows}_{En}$, and $r^{(0)\rightleftarrows}_{En}$ are calculated by applying the boundary conditions 
\be 
\psi^{(0)}_{En}(+0)=\psi^{(0)}_{En}(-0),\ \ \ 
\frac{d\psi^{(0)}_{En}(+0)}{dx}=\frac{d\psi^{(0)}_{En}(-0)}{dx}.
\ee 
They are
\be 
t^{(0)\rightrightarrows}_{En}=\frac 2{1+\frac{Q(E-E_Wn^2-V_R)}{Q(E-E_Wn^2-V_L)} },\ \ 
r^{(0)\leftrightarrows}_{En}=\frac {1-\frac{Q(E-\epsilon_n-V_R)}{Q(E-E_Wn^2-V_L)} }{1+\frac{Q(E-E_Wn^2-V_R)}{Q(E-E_Wn^2-V_L)} },
\label{tr->}
\ee
\be 
t^{(0)\leftleftarrows}_{En}=\frac 2{1+\frac{Q(E-E_Wn^2-V_L)}{Q(E-E_Wn^2-V_R)} },\ \ 
r^{(0)\rightleftarrows}_{En}=\frac {1-\frac{Q(E-E_Wn^2-V_L)}{Q(E-E_Wn^2-V_R)} }{1+\frac{Q(E-E_Wn^2-V_L)}{Q(E-E_Wn^2-V_R)} }.
\label{tr<-}
\ee
Here the lower and upper arrows in the superscripts of the transmission and reflection amplitudes indicate the direction of motion of the electron wave incident on the potential step and going from it, see Figure \ref{fig:DoubleArrows}, illustrating the definitions (\ref{tr->})--(\ref{R0}). The results for the energies $V_L<E-E_Wn^2<V_R$ lying above $V_L$ but below $V_R$ are obtained from (\ref{PsiRight})--(\ref{PsiLeft}) and (\ref{tr->})--(\ref{tr<-}) with the help of the analytical continuation (\ref{Qof-E}).

\begin{figure}
\includegraphics[width=0.49\textwidth]{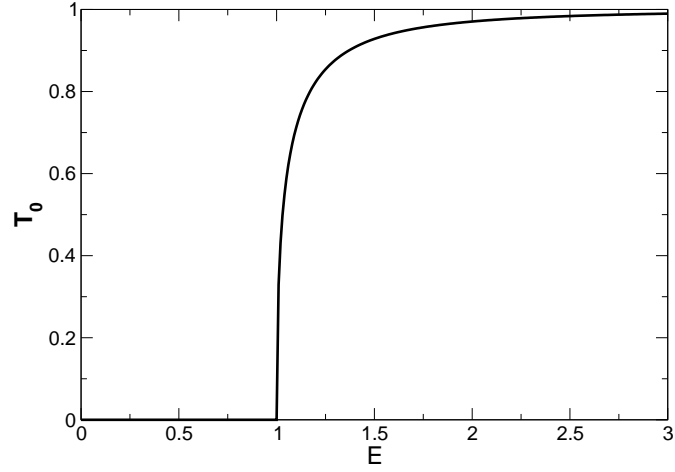}
\caption{\label{fig:TR0} The function $\mathsf{T}_{0}({\cal E})$ defined by  Eq. (\ref{calT0}). }
\end{figure}

The zeroth-order transmission and reflection coefficients are then calculated as the ratios of the particle flows of the transmitted and reflected waves to the particle flow of the incident wave,
\be 
T_{En}^{(0)\rightrightarrows} =\frac{Q(E-E_Wn^2-V_R)}{Q(E-E_Wn^2-V_L)} \left|t^{(0)\rightrightarrows}_{En}\right|^2, \ \ \ R^{(0)\leftrightarrows}_{En}=\left|r^{(0)\leftrightarrows}_{En}\right|^2, \label{T0}
\ee 
\be 
T_{En}^{(0)\leftleftarrows} =\frac{Q(E-E_Wn^2-V_L)}{Q(E-E_Wn^2-V_R)} \left|t^{(0)\leftleftarrows}_{En}\right|^2, \ \ \ R^{(0)\rightleftarrows}_{En}=\left|r^{(0)\rightleftarrows}_{En}\right|^2. 
\label{R0}
\ee 
The result can be written in the form
\be 
T_{En}^{(0)\rightrightarrows} =T_{En}^{(0)\leftleftarrows} \equiv T_{En}^{(0)} 
=\mathsf{T}_{0}\left(\frac{E-E_Wn^2-V_L}{V_B}\right) , 
\label{Tcoeff0}
\ee 
\be 
R^{(0)\leftrightarrows}_{En}=R^{(0)\rightleftarrows}_{En}\equiv R^{(0)}_{En}=1-T_{En}^{(0)} ,
\label{Rcoeff0}
\ee
where $V_B=V_R-V_L$ is the height of the potential step and $\mathsf{T}_{0}$ designates the dimensionless function 
\be 
\mathsf{T}_{0}({\cal E}) =\Theta({\cal E}-1)
\frac {4\sqrt{{\cal E}({\cal E}-1)}} {\left|\sqrt{{\cal E}}+\sqrt{{\cal E}-1}\right|^2 };
\label{calT0}
\ee
it is nothing but the transmission coefficient of a one-dimensional step potential (see e.g. Ref. \cite{Landau3}, problem 1 to \S 25), where the energy is counted from $V_L$ and is measured in units of the step height $V_B$. The energy dependence of the function $\mathsf{T}_{0}$ is shown in Figure \ref{fig:TR0}. 

\subsection{Quantum conductance \label{sec:quant-conductance}}

\begin{figure}
\includegraphics[width=0.49\textwidth]{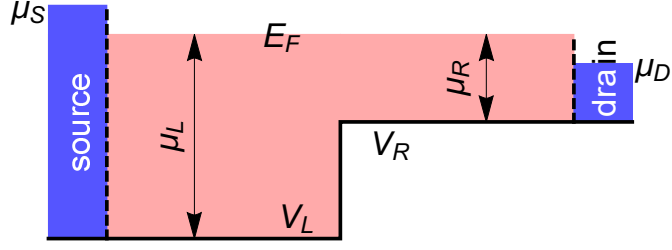}
\caption{\label{fig:SDcircuit} The source-drain circuit. $\mu_S$ and $\mu_D$ are the source and drain chemical potentials; their difference is generated by the source - drain voltage $U_{sd}$. $\mu_L$ and $\mu_R$ are the chemical potentials under the left and right gates produced by the gate voltages $U_L$ and $U_R$.}
\end{figure}

Using results obtained in the previous Section we can now calculate the quantum conductance of the region $|x|\lesssim b\ll l_{\mathrm{mfp}}$ containing the potential step $V_0(x)$. Assume that different voltages $U_L$ and $U_R$ are applied to the left and right gates, and a dc voltage $U_{sd}$ is applied between the source and drain contacts, Figure \ref{fig:SDcircuit}. The voltage $U_{sd}$ generates a difference of the source ($\mu_S$) and drain ($\mu_D$) chemical potentials, $\mu_S-\mu_D=-eU_{sd}$. Particles flowing to the right, in the source to drain direction, create an electric current 
\ba 
I_x^\rightarrow=
-e\sum_n\frac {g_s}{L}\sum_{k_x>0}v_xT_{En}^{(0)\rightrightarrows}
F(E-\mu_S,T)\Big(1-F(E-\mu_D,T)\Big),
\label{I0->}
\ea
where the wavevector $k_x$ is related to the total energy $E$ as 
\be 
E=V_L+E_Wn^2+\frac{\hbar^2k_x^2}{2m}.
\ee
Electrons flowing to the left, in the drain to source direction, produce an electric current 
\ba 
I_x^\leftarrow=-e\sum_n\frac {g_s}{L}\sum_{k_x<0}v_x
T_{En}^{(0)\leftleftarrows}
F(E-\mu_D,T)\Big(1-F(E-\mu_S,T)\Big),
\label{I0<-}
\ea
where the wavevector $k_x$ is related to the energy $E$ by the formula
\be 
E=V_R+E_Wn^2+\frac{\hbar^2k_x^2}{2m}.
\ee
In both formulas (\ref{I0->}) and (\ref{I0<-}), the velocity $v_x$ equals $v_x=\hbar k_x/m$, and the transmission probabilities $T_{En}^{(0)\rightrightarrows}=T_{En}^{(0)\leftleftarrows}$ are determined by Eqs. (\ref{Tcoeff0}), (\ref{calT0}). The factors containing the Fermi distribution function 
\be 
F(E,T)=\left[1+\exp\left(\frac{E}{T}\right)\right]^{-1}
\label{FermiFunction}
\ee 
take into account the occupation probabilities of quantum states and the Pauli principle (we omit the Boltzmann constant $k_B$ here and below). 

\begin{figure}
\includegraphics[width=0.49\textwidth]{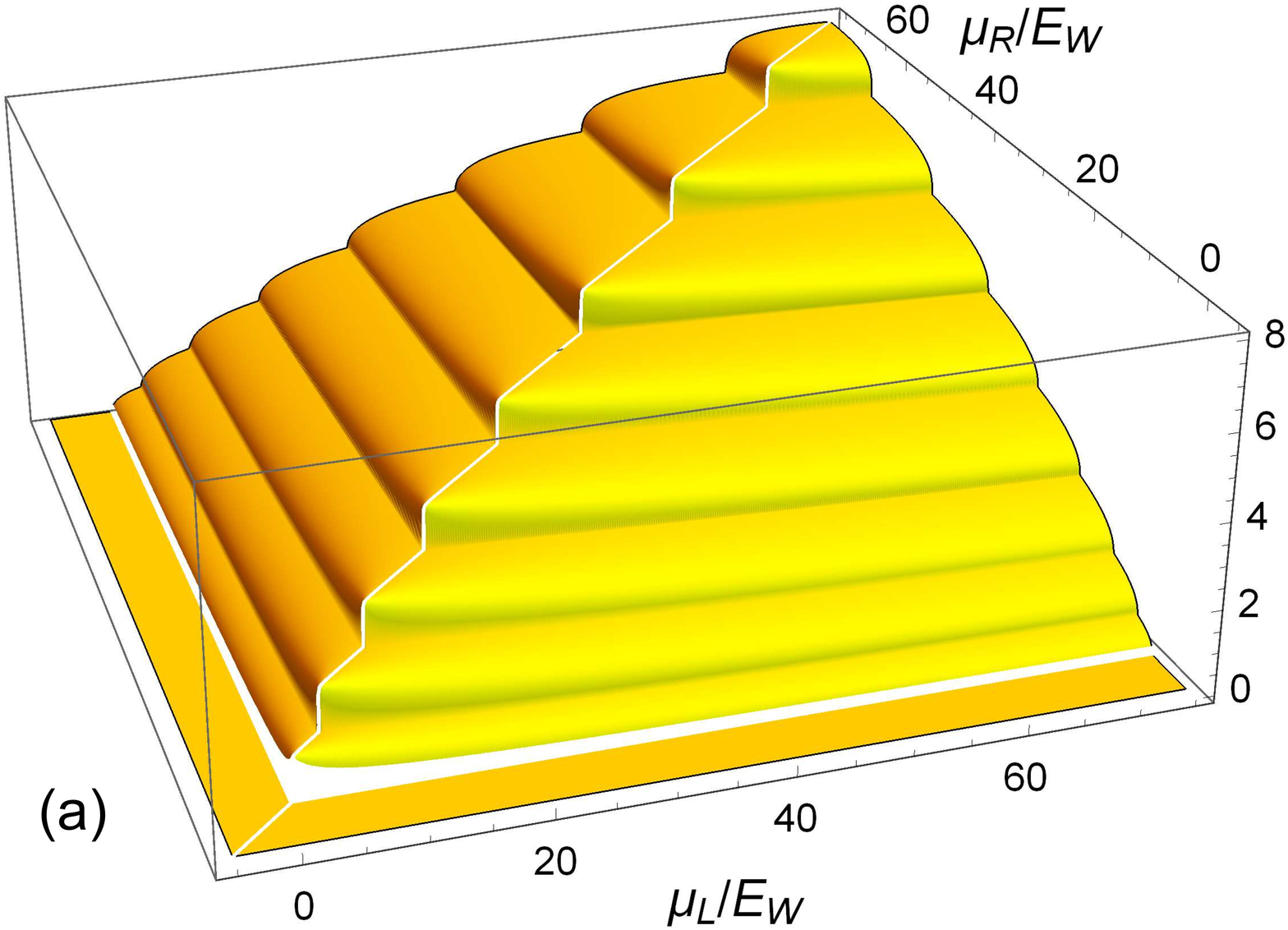}
\includegraphics[width=0.49\textwidth]{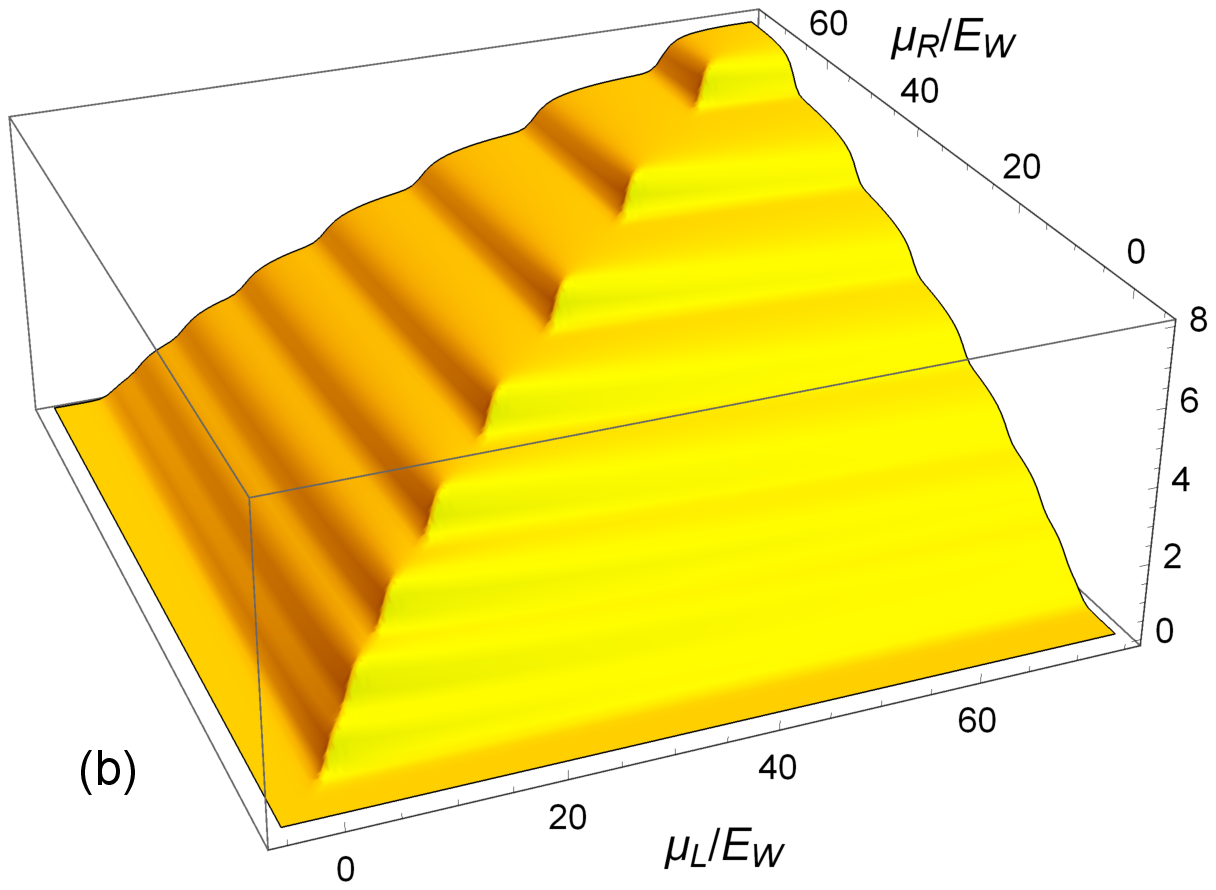}
\includegraphics[width=0.49\textwidth]{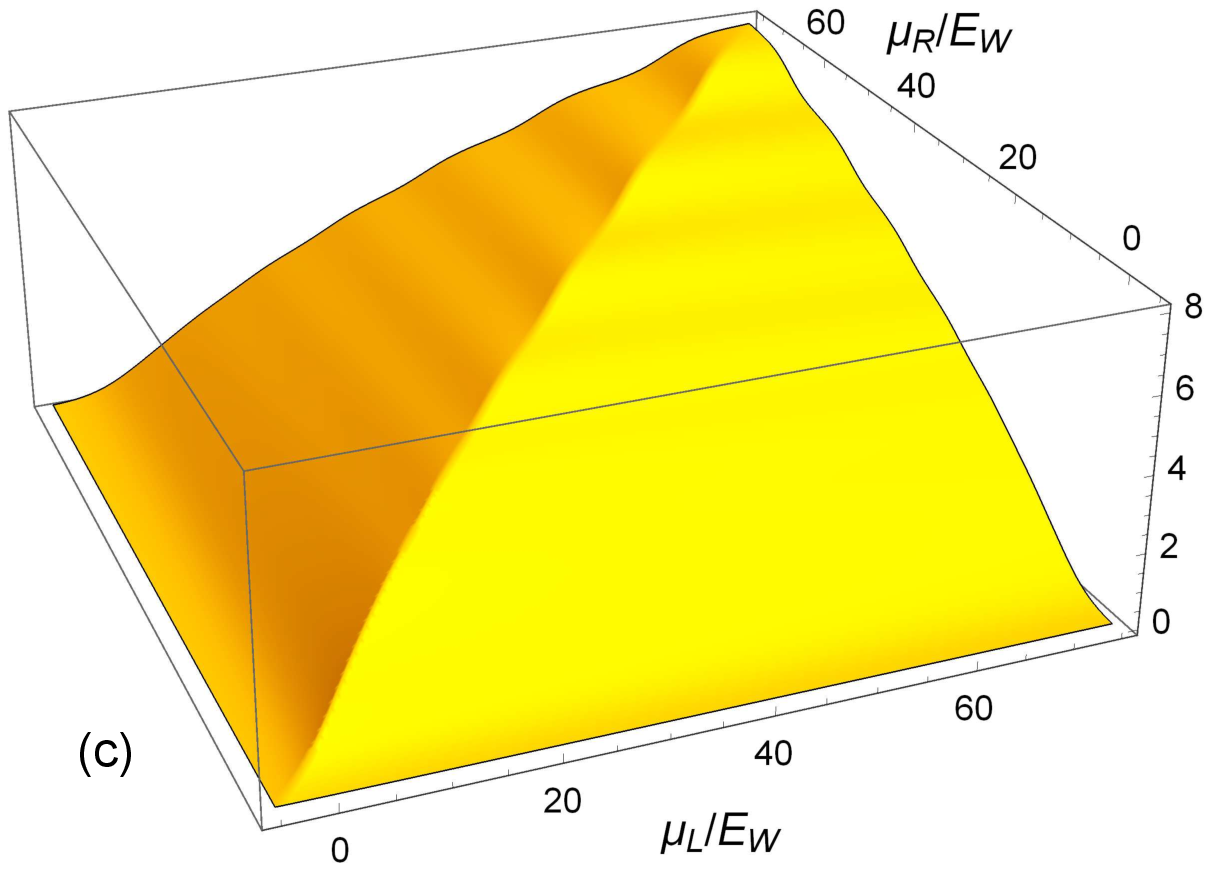}
\includegraphics[width=0.49\textwidth]{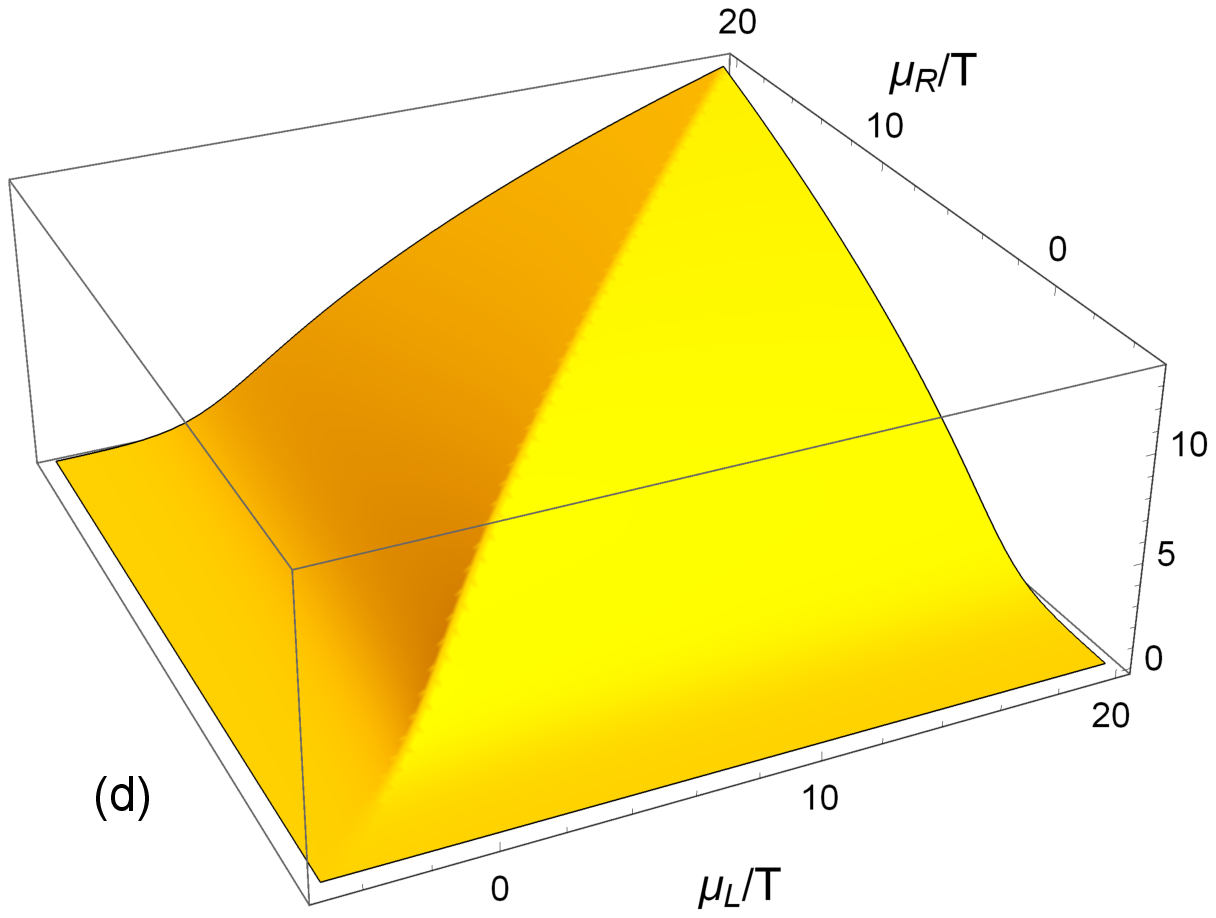}
\caption{\label{fig:QuantCond} The quantum conductance $\sigma_q$, Eq. (\ref{quant-conduct}), in units $e^2/\pi\hbar$, at different values of the parameter $T/E_W$: (a) $T/E_W=0$, (b) $T/E_W=0.5$, (c) $T/E_W=2$, and (d) $T/E_W=10$. Notice that in the classical (high temperature) regime $T\gg E_W$ the chemical potentials are normalized to the larger energy scale $T$.}
\end{figure}

The total direct current in the system is given by the sum of the two currents (\ref{I0->}) and (\ref{I0<-}) and can be presented in the form
\be 
I^{(0)}=-\frac{e}{\pi\hbar} \sum_{n=1}^\infty 
\int_{-\infty}^\infty  dE 
\left[
T_{En}^{(0)\rightrightarrows} F(E-\mu_S,T)\Big(1-F(E-\mu_D,T)\Big)- 
T_{En}^{(0)\leftleftarrows} F(E-\mu_D,T)\Big(1-F(E-\mu_S,T)\Big)
\right].
\label{current0order-general}
\ee
In the linear response regime $eU_{sd}\ll T$, we get from here, after some algebra, $I^{(0)}=\sigma_qU_{\rm sd}$, with the quantum conductance
\be 
\sigma_q(\mu_L,\mu_R,T,E_W)=
\frac {e^2 }{\pi\hbar } \frac{E_W}{T}\int^{\min\{\xi_L,\xi_R\}}_{-\infty} dX
\frac {\sqrt{(\xi_L-X)(\xi_R-X)}} {\left|\sqrt{\xi_L-X}+\sqrt{\xi_R-X}\right|^2 }
\sum_{n=1}^\infty 
\frac { 1 }
{\cosh^2\left(\frac{X-n^2}{2(T/E_W)}\right)},
\label{quant-conduct}
\ee 
where $\xi_{L,R}=\mu_{L,R}/{E_W}$. Figure \ref{fig:QuantCond} shows the quantum conductance $\sigma_q(\mu_L,\mu_R,T,E_W)$, in units $e^2/\pi\hbar$, as a function of the dimensionless chemical potentials $\mu_L$ and $\mu_R$ at different values of the parameter $E_W/T$. Figure \ref{fig:QuantCond}(a) illustrates $\sigma_q$ at zero temperature $T=0$. In this case the result (\ref{quant-conduct}) can be simplified:
\ba 
\sigma_q(\mu_L,\mu_R,T=0,E_W)=
\frac {e^2 }{\pi\hbar } \sum_{n=1}^{\sqrt{\min\{\xi_L,\xi_R\}} } 
\frac {4\sqrt{(\xi_L-n^2)(\xi_R-n^2)}} {\left|\sqrt{\xi_L-n^2}+\sqrt{\xi_R-n^2}\right|^2 }.
\label{quant-conduct-T=0}
\ea 
The upper limit in the sum in (\ref{quant-conduct-T=0}) is assumed to be larger then 1, $\min\{\xi_L,\xi_R\}\ge 1$; otherwise $\sigma_q=0$. On the diagonal, when $\mu_L=\mu_R$ and the potential step is absent, the result (\ref{quant-conduct-T=0}) reproduces the conductance quantization effect, derived and experimentally observed in narrow ballistic 2D channels in Refs. \cite{Wees88,Wharam88}. The formula (\ref{quant-conduct-T=0} generalizes the result of \cite{Wees88,Wharam88} to the case when inside the channel electrons meet a potential step on their way from source to drain. In Figures \ref{fig:QuantCond}(b) and (c), plotted for $T/E_W=0.5$ and $2$, we show how the quantized conductance steps are smoothed out when the temperature grows. They are seen quite well at $T/E_W=0.5$, Figure \ref{fig:QuantCond}(b), but become almost invisible when this parameters equals 2, Figure \ref{fig:QuantCond}(c). 

The quantum limit $T/E_W\lesssim 1$ corresponds to the case of very narrow 2D channels and low temperatures. For example, in GaAs quantum wells with the 2D channel width $W=0.2$ $\mu$m, the condition $T/E_W<1$ is satisfied at temperatures $T<1.6$ K. A more practical case of relatively wide ($W\gtrsim 1$ $\mu$m) channels and not so low temperatures (e.g. $T\gtrsim 4$ K) corresponds to the classical limit. This case, that was realized in Ref. \cite{Michailow22} and corresponds to the condition $T/E_W\gg 1$, is illustrated in Figure \ref{fig:QuantCond}(d). The dependence of $\sigma_q$ on the chemical potentials is smooth in this limit. Notice that in Figure \ref{fig:QuantCond}(d) we normalize the arguments of the function $\sigma_q$, $\mu_L/T$ and $\mu_R/T$, to the larger energy scale $T$.  

\section{First-order approximation\label{sec:1storder}}

\subsection{Transmission and reflection of photo-excited electrons\label{sec:TR+-}}

In the first-order perturbation theory the equation to be solved has the form
\be 
i\hbar \frac{\p \Psi^{(1)}}{\p t}=\hat H_0\Psi^{(1)}+\hat H_1\Psi^{(0)},
\ee
so that we get an inhomogeneous Schr\"odinger equation
\be 
i\hbar \frac{\p \Psi^{(1)}}{\p t}+\frac{\hbar^2}{2m}\frac{\p^2 \Psi^{(1)}}{\p x^2}+\frac{\hbar^2}{2m}\frac{\p^2 \Psi^{(1)}}{\p y^2}-V_0(x)\Psi^{(1)}
=\frac {e\Delta\Phi_{ac}}4
\left(e^{-i(E+\hbar\omega)t/\hbar}+e^{-i(E-\hbar\omega)t/\hbar}\right)
\sign(x) \sin\frac{\pi n y}{W}\psi_{En}^{(0)}(x). 
\label{SE1storder}
\ee
Two time-dependent terms in the right-hand side correspond to the absorption (energy $E+\hbar\omega$) and emission (energy $E-\hbar\omega$) of a photon. Since the second-order differential equation (\ref{SE1storder}) is linear, the response to each of the time-dependent exponents can be searched for separately. By substituting the first-order wave function in the form
\be 
\Psi^{(1)}(x,y,t)=e^{-i(E\pm\hbar\omega)t/\hbar}\sin\frac{\pi n y}{W}\psi_{En}^{(1)}(x) 
\ee
and canceling the time- and $y$-dependent factors, we get the differential equation for $\psi_{En}^{(1)}(x)$:
\ba 
\frac{\hbar^2}{2m}\frac{\p^2 \psi_{En}^{(1)}(x)}{\p x^2}+
\left[(E\pm\hbar\omega) 
-E_W n^2
 -V_L-(V_R-V_L)\Theta(x) \right]\psi_{En}^{(1)}(x)=
\frac 14e\Delta\Phi_{ac}\sign(x) \psi_{En}^{(0)}(x) .\label{SE1storderXdep}
\ea
This equation should be solved for all energies $E-E_W n^2>V_L$ and for both directions of the incident waves.

Consider first the case of the over-barrier electrons with energies $\tilde E\equiv E-E_W n^2 >V_R$ running to the right. Then equation (\ref{SE1storderXdep}) takes the form
\be 
\frac{\hbar^2}{2m}\frac{\p^2 \psi_{En}^{(1)}(x)}{\p x^2}+
\left(\tilde E\pm\hbar\omega -V_L\right)\psi_{En}^{(1)}(x)=
-\frac {1 }4 e\Delta\Phi_{ac}\left(e^{iQ(\tilde E-V_L)x}+r_{En}^{(0)\leftrightarrows}e^{-iQ(\tilde E-V_L)x}\right)
\label{eq1st_x<0}
\ee
at $x<0$, and 
\be 
\frac{\hbar^2}{2m}\frac{\p^2 \psi_{En}^{(1)}(x)}{\p x^2}+
\left(\tilde E \pm\hbar\omega -V_R \right)\psi_{En}^{(1)}(x)=
\frac 14e\Delta\Phi_{ac} t_{En}^{(0)\rightrightarrows}e^{iQ(\tilde E-V_R)x}
\label{eq1st_x>0}
\ee
at $x>0$, with the coefficients $t_{En}^{(0)\rightrightarrows}$ and $r_{En}^{(0)\leftrightarrows}$ given by Eqs. (\ref{tr->}). Solutions of these equations are written as 
\be 
\psi_{En}^{(1)}(x)=
A_{En}^\pm e^{-iQ(\tilde E\pm\hbar\omega-V_L)x} - \frac {e\Delta\Phi_{ac}}{4(\pm\hbar\omega)}\left(e^{iQ(\tilde E-V_L)x}+r_{En}^{(0)}e^{-iQ(\tilde E-V_L)x}\right), \ \ x<0,
\label{sol1st_x<0}
\ee
\be 
\psi_{En}^{(1)}(x)=
B_{En}^\pm e^{iQ(\tilde E\pm\hbar\omega-V_R)x} + \frac {e\Delta\Phi_{ac}}{4(\pm\hbar\omega)} t_{En}^{(0)}e^{iQ(\tilde E-V_R) x},\  \ x>0,
\label{sol1st_x>0}
\ee
where the boundary conditions at $x=\pm \infty$, corresponding to the absence of the waves with the energy $E\pm\hbar\omega$ coming from infinity, are already taken into account. The coefficients $A_{En}^\pm$ and $B_{En}^\pm$ are determined from the boundary conditions at $x=0$ (the continuity of the wave function and its derivative). They are
\be 
A_{En}^\pm=
- \frac {e\Delta\Phi_{ac}}{(\pm\hbar\omega)}
\frac {Q(\tilde E-V_L) \Big(Q(\tilde E-V_R) -Q(\tilde E\pm\hbar\omega-V_R) \Big)  }
{\Big(Q(\tilde E-V_L)+Q(\tilde E-V_R) \Big)\Big(Q(\tilde E\pm\hbar\omega-V_L) +Q(\tilde E\pm\hbar\omega-V_R)\Big)},
\ee
\be
B_{En}^\pm
=
-\frac {e\Delta\Phi_{ac}}{(\pm\hbar\omega)} \frac{Q(\tilde E-V_L)
\Big(Q(\tilde E-V_R)   + Q(\tilde E\pm\hbar\omega-V_L)\Big)}
{\Big(Q(\tilde E-V_L)+Q(\tilde E-V_R) \Big)\Big(Q(\tilde E\pm\hbar\omega-V_L) +Q(\tilde E\pm\hbar\omega-V_R)\Big)}.
\ee

Now we can calculate the probability of an electron wave, incident on the potential step from the left, to absorb or emit a photon and to continue to move in the same direction. The particle flow of the incident wave is then proportional to $Q(E-E_Wn^2-V_L)$. The particle flow of the transmitted wave, after absorption or emission of a photon, is proportional to
\be 
\Theta(E\pm\hbar\omega-E_Wn^2-V_R)
Q(E\pm\hbar\omega-E_Wn^2-V_R)|B_{En}^\pm|^2.
\ee
The required ``transmission coefficient'' of the electron, that absorbed or emitted a photon, in the first-order perturbation theory is then
\be
T_{En}^{\pm\rightrightarrows}=\Theta(E\pm\hbar\omega-E_Wn^2-V_R)
\frac{Q(E\pm\hbar\omega-E_Wn^2-V_R)}{Q(E-E_Wn^2-V_L)}|B_{En}^\pm|^2.
\ee
This formula can be transformed into the following compact form
\be
T_{En}^{\pm\rightrightarrows}=
\left(\frac {e\Delta\Phi_{ac}}{\hbar\omega}\right)^2 
\mathsf{T}_{\pm}^{\rightrightarrows}\left(\frac{E-E_Wn^2-V_L}{V_B},\frac{\hbar\omega}{V_B}\right),
\label{T->->}
\ee
where 
\be
\mathsf{T}_\pm^{\rightrightarrows}\left({\cal E}, \Omega\right)=\Theta({\cal E})\Theta({\cal E}\pm\Omega-1)
\frac{\sqrt{{\cal E}}\sqrt{{\cal E}\pm\Omega-1}
\left|\sqrt{{\cal E}-1}+\sqrt{{\cal E}\pm\Omega}\right|^2}
{\left|\sqrt{{\cal E}}+\sqrt{{\cal E}-1} \right|^2 \left|\sqrt{{\cal E}\pm\Omega} +\sqrt{{\cal E}\pm\Omega-1}\right|^2}
\label{calT->->}
\ee
is a function of only two dimensionless parameters, 
\be 
{\cal E}=\frac{E-E_Wn^2-V_L}{V_B}\ \textrm{and }\Omega=\frac{\hbar\omega}{V_B}:
\ee
the energy of electrons counted from the left conduction band bottom and the photon energy, both measured in units of the potential step energy $V_B$. 

In a similar manner we calculate the ``reflection coefficient'' of electrons $R_{En}^{\pm\leftrightarrows}$ after absorption or emission of a photon, as well as the $T$ and $R$ coefficients for the particles incident on the potential step from the right. The final results are formulated as follows.

\begin{enumerate}
\item Electrons \textit{running to the right} may absorb or emit a photon and continue to move in the same direction (\textit{to the right}). The ``transmission'' coefficient corresponding to this process is determined by Eqs. (\ref{T->->})--(\ref{calT->->}). 

\item Electrons \textit{running to the right} may absorb or emit a photon and continue to move in the opposite direction (\textit{to the left}). The ``reflection'' coefficient corresponding to this process equals 
\be
R_{E,n}^{\pm\leftrightarrows}=
\left(\frac {e\Delta\Phi_{ac}}{\hbar\omega}\right)^2 
\mathsf{R}_{\pm}^{\leftrightarrows}\left(\frac{E-E_Wn^2-V_L}{V_B},\frac{\hbar\omega}{V_B}\right)
\label{R-><-}
\ee
where 
\be
 \mathsf{R}_{\pm}^{\leftrightarrows}\left({\cal E},\Omega\right)=\Theta({\cal E})\Theta({\cal E}\pm\Omega)
\frac{\sqrt{{\cal E}}\sqrt{{\cal E}\pm\Omega}
\left|\sqrt{{\cal E}-1}-\sqrt{{\cal E}\pm\Omega-1}\right|^2}
{\left|\sqrt{{\cal E}}+\sqrt{{\cal E}-1} \right|^2 \left|\sqrt{{\cal E}\pm\Omega} +\sqrt{{\cal E}\pm\Omega-1}\right|^2}.
\label{calR-><-}
\ee

\item Electrons \textit{running to the left} may absorb or emit a photon and continue to move in the same direction (\textit{to the left}). The ``transmission'' coefficient corresponding to this process equals 
\be
T_{En}^{\pm\leftleftarrows}=
\left(\frac {e\Delta\Phi_{ac}}{\hbar\omega}\right)^2 
\mathsf{T}_{\pm}^{\leftleftarrows}\left(\frac{E-E_Wn^2-V_L}{V_B},\frac{\hbar\omega}{V_B}\right)
\label{T<-<-}
\ee
where 
\be
\mathsf{T}_\pm^{\leftleftarrows}\left({\cal E},\Omega\right)=\Theta({\cal E}-1)\Theta({\cal E}\pm\Omega)
\frac{\sqrt{{\cal E}-1}\sqrt{{\cal E}\pm\Omega}
\left|\sqrt{{\cal E}}+\sqrt{{\cal E}\pm\Omega-1}\right|^2}
{\left|\sqrt{{\cal E}}+\sqrt{{\cal E}-1} \right|^2 \left|\sqrt{{\cal E}\pm\Omega} +\sqrt{{\cal E}\pm\Omega-1}\right|^2}.
\label{calT<-<-}
\ee

\item Electrons \textit{running to the left} may absorb or emit a photon and continue to move in the opposite direction (\textit{to the right}). The ``reflection'' coefficient corresponding to this process equals 
\be
R_{E,n}^{\pm\rightleftarrows}=
\left(\frac {e\Delta\Phi_{ac}}{\hbar\omega}\right)^2 
\mathsf{R}_{\pm}^{\rightleftarrows}\left(\frac{E-E_Wn^2-V_L}{V_B},\frac{\hbar\omega}{V_B}\right)
\label{R<-->}
\ee
where 
\be
\mathsf{R}_{\pm}^{\rightleftarrows}\left({\cal E},\Omega\right)=\Theta({\cal E}-1)\Theta({\cal E}\pm\Omega-1)
\frac{\sqrt{{\cal E}-1}\sqrt{{\cal E}\pm\Omega-1}
\left|\sqrt{{\cal E}}-\sqrt{{\cal E}\pm\Omega}\right|^2}
{\left|\sqrt{{\cal E}}+\sqrt{{\cal E}-1} \right|^2 \left|\sqrt{{\cal E}\pm\Omega} +\sqrt{{\cal E}\pm\Omega-1}\right|^2}.
\label{calR<-->}
\ee
\end{enumerate}
The formulas (\ref{T->->})--(\ref{calR<-->}) are valid at all energies (below and above the barrier). The formulas for $T_{En}^{\pm\leftleftarrows}$ and $R_{En}^{\pm\rightleftarrows}$ can be obtained from $T_{En}^{\pm\rightrightarrows}$ and $R_{En}^{\pm\leftrightarrows}$ by exchanging $V_L\leftrightarrow V_R$. The prefactor $\alpha=\left(e\Delta\Phi_{ac}/\hbar\omega\right)^2$ in all formulas is the perturbation theory parameter, determined by the ratio of the ac potential difference between the antenna wings to the photon energy. It should be smaller than one for the  theory to be valid; if $\alpha\gtrsim 1$ higher orders of the perturbation theory have to be taken into account. In the experiment \cite{Michailow22} $\alpha$ was about 0.1. It is important to emphasize that, while the transmission coefficients $T_{En}^{(0)\rightrightarrows}$ and $T_{En}^{(0)\leftleftarrows}$ without irradiation are equal, Eq. (\ref{Tcoeff0}), the coefficients $T_{En}^{\pm\rightrightarrows}$ and $T_{En}^{\pm\leftleftarrows}$ under irradiation are substantially different, see Eqs. (\ref{T->->}) and (\ref{T<-<-}). The same is valid, of course, for the $R$ coefficients, too.

As was mentioned in the Introduction, in this paper we will analyze only the photocurrent response of the structure at zero temperature. In this case, the processes of photon emission do not contribute to the photocurrent due to the Pauli principle, see Section \ref{sec:photocurrent} for details. Therefore, in what follows, we will analyze and discuss only the coefficients $T_{En}^{+\rightrightarrows}$, $T_{En}^{+\leftleftarrows}$, $R_{E,n}^{+\leftrightarrows}$, and $R_{E,n}^{+\rightleftarrows}$, which describe the first order transmission and reflection processes after absorption of THz photons.

\begin{figure}
\includegraphics[width=0.49\textwidth]{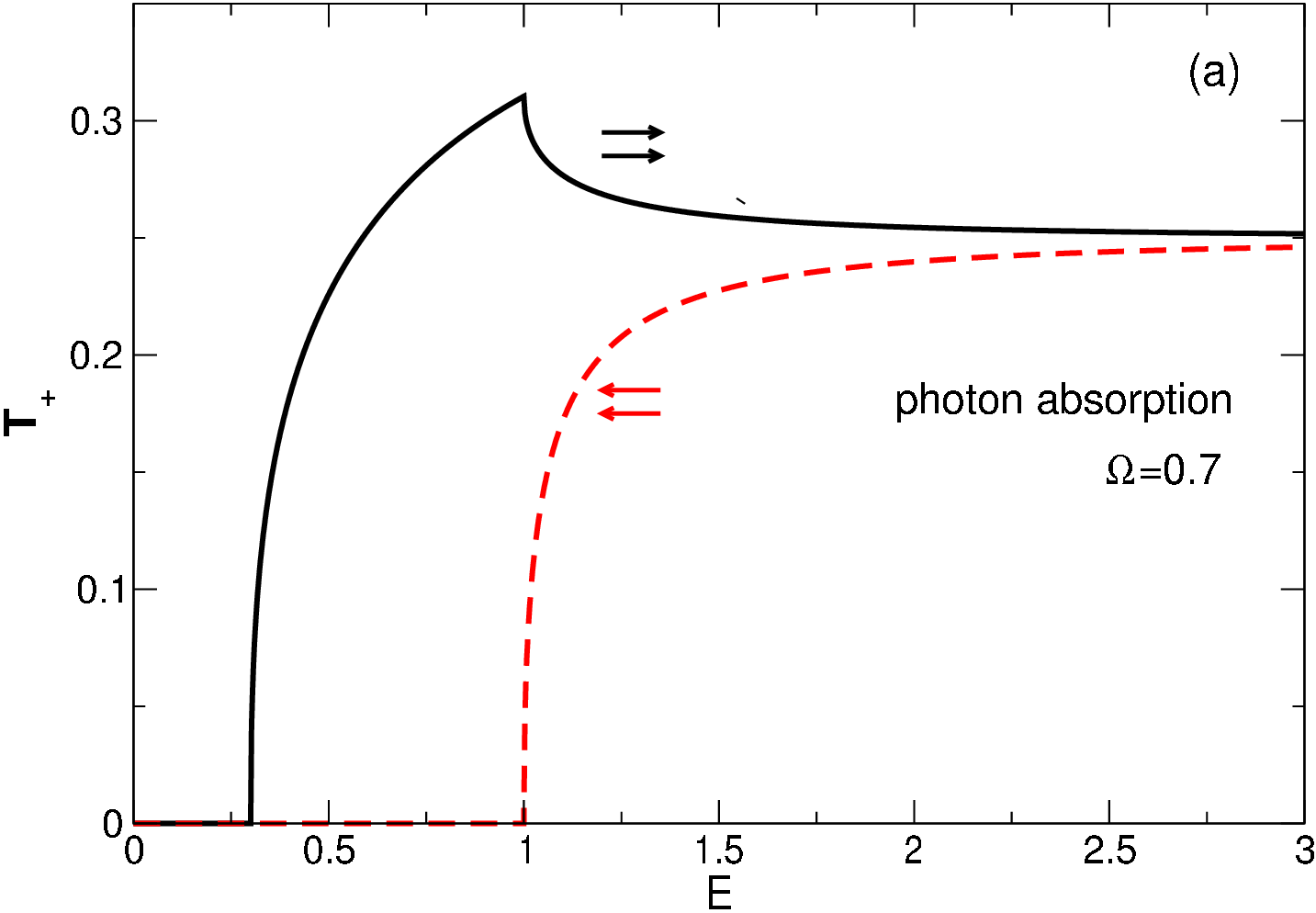}
\includegraphics[width=0.49\textwidth]{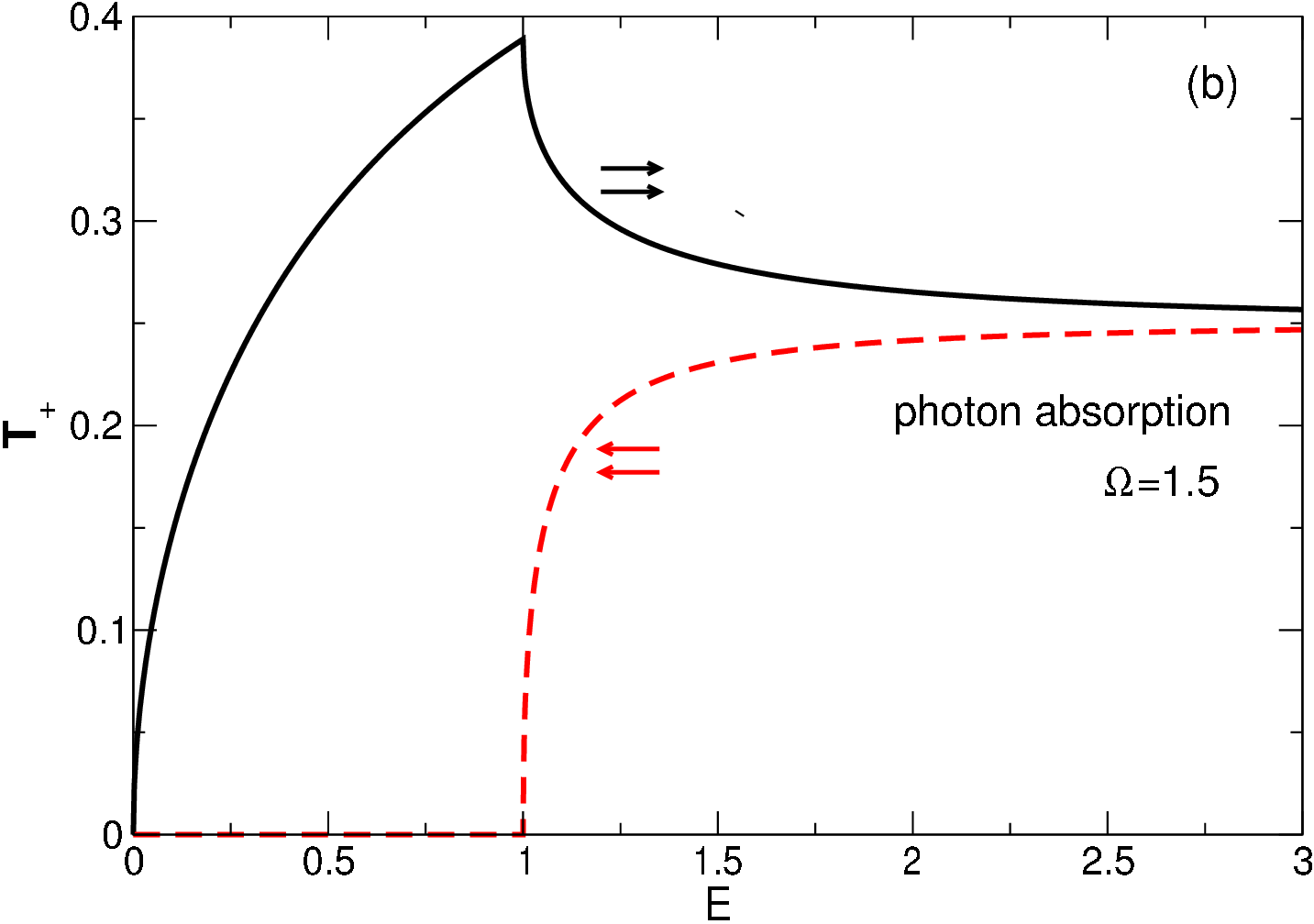}
\caption{\label{fig:T+} The functions $\mathsf{T}_{+}^{\rightrightarrows}({\cal E},\Omega)$ and $\mathsf{T}_{+}^{\leftleftarrows}({\cal E},\Omega)$ defined by equations (\ref{calT->->}) and (\ref{calT<-<-}) for (a) $\Omega<1$ and (b) $\Omega>1$.}
\end{figure}

Let us consider the energy and frequency dependencies of these coefficients. Figure \ref{fig:T+} illustrates the energy dependence of the dimensionless transmission functions $\mathsf{T}_{+}^{\rightrightarrows}({\cal E},\Omega)$ and $\mathsf{T}_{+}^{\leftleftarrows}({\cal E},\Omega)$ at a photon energy smaller (Fig. \ref{fig:T+}(a)) and larger (Fig. \ref{fig:T+}(b)) than the potential step height. One sees that the transmission coefficient of particles running onto the step ($\mathsf{T}_{+}^{\rightrightarrows}$) is always larger than that of those running from the step ($\mathsf{T}_{+}^{\leftleftarrows}$). At the energies below the step height the function $\mathsf{T}_{+}^{\leftleftarrows}({\cal E},\Omega)$ vanishes, while the function $\mathsf{T}_{+}^{\rightrightarrows}({\cal E},\Omega)$ is finite either at all energies, if $\hbar\omega>V_B$, or at energies $E>V_B-\hbar\omega$, if $\hbar\omega<V_B$. At large energies $E\gg V_B$ both functions $\mathsf{T}_{+}^{\rightrightarrows}({\cal E},\Omega)$ and $\mathsf{T}_{+}^{\leftleftarrows}({\cal E},\Omega)$ tend toward the value $1/4$, from above and from below, respectively. Since the photocurrent (discussed below in Section \ref{sec:photocurrent}) is determined by the difference $\mathsf{T}_{+}^{\rightrightarrows}({\cal E},\Omega)-\mathsf{T}_{+}^{\leftleftarrows}({\cal E},\Omega)$, the photoexcited electrons always move onto the step, i.e., from the area with a higher electron density to the area with a lower electron density.

\begin{figure}
\includegraphics[width=0.49\textwidth]{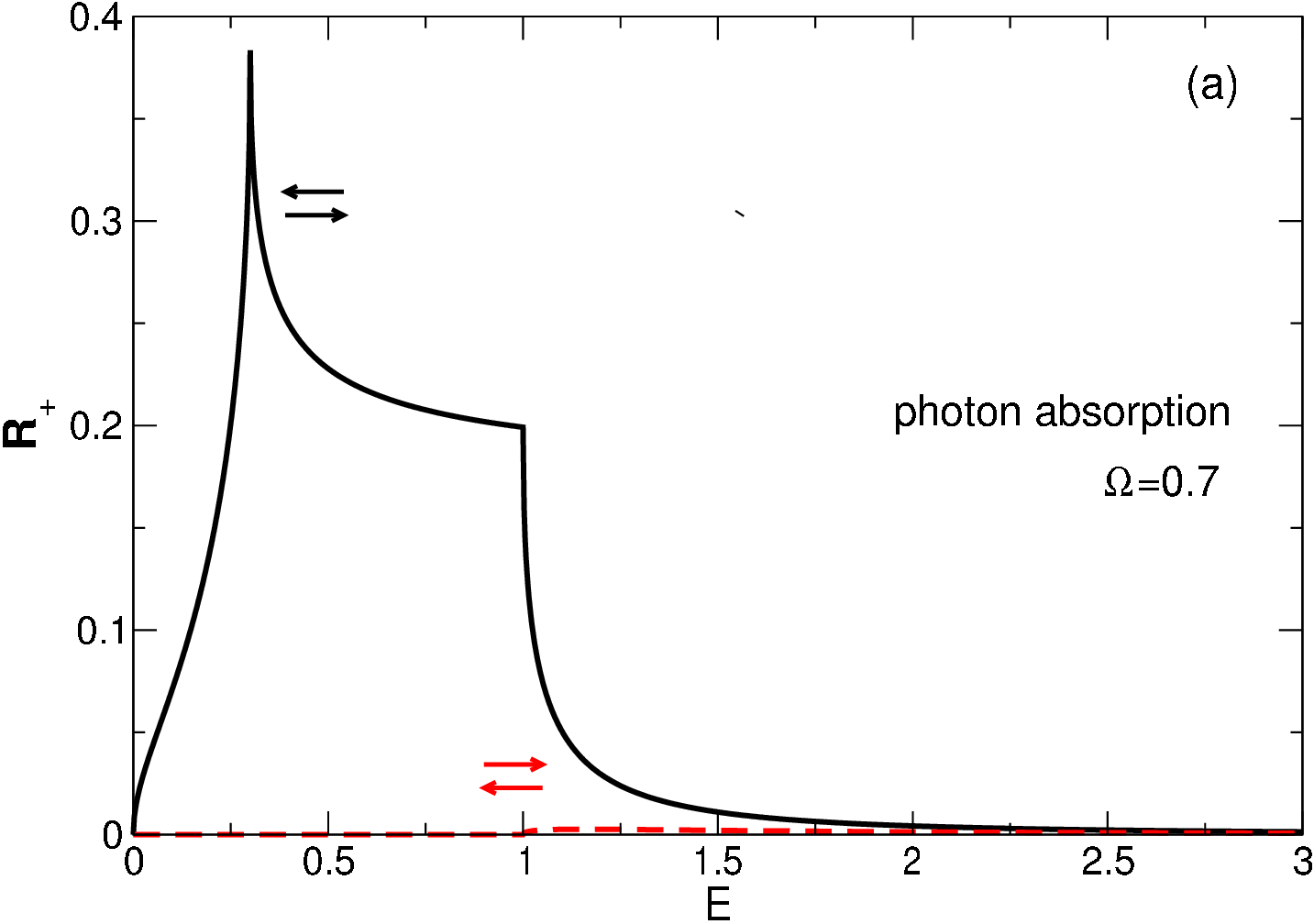}
\includegraphics[width=0.49\textwidth]{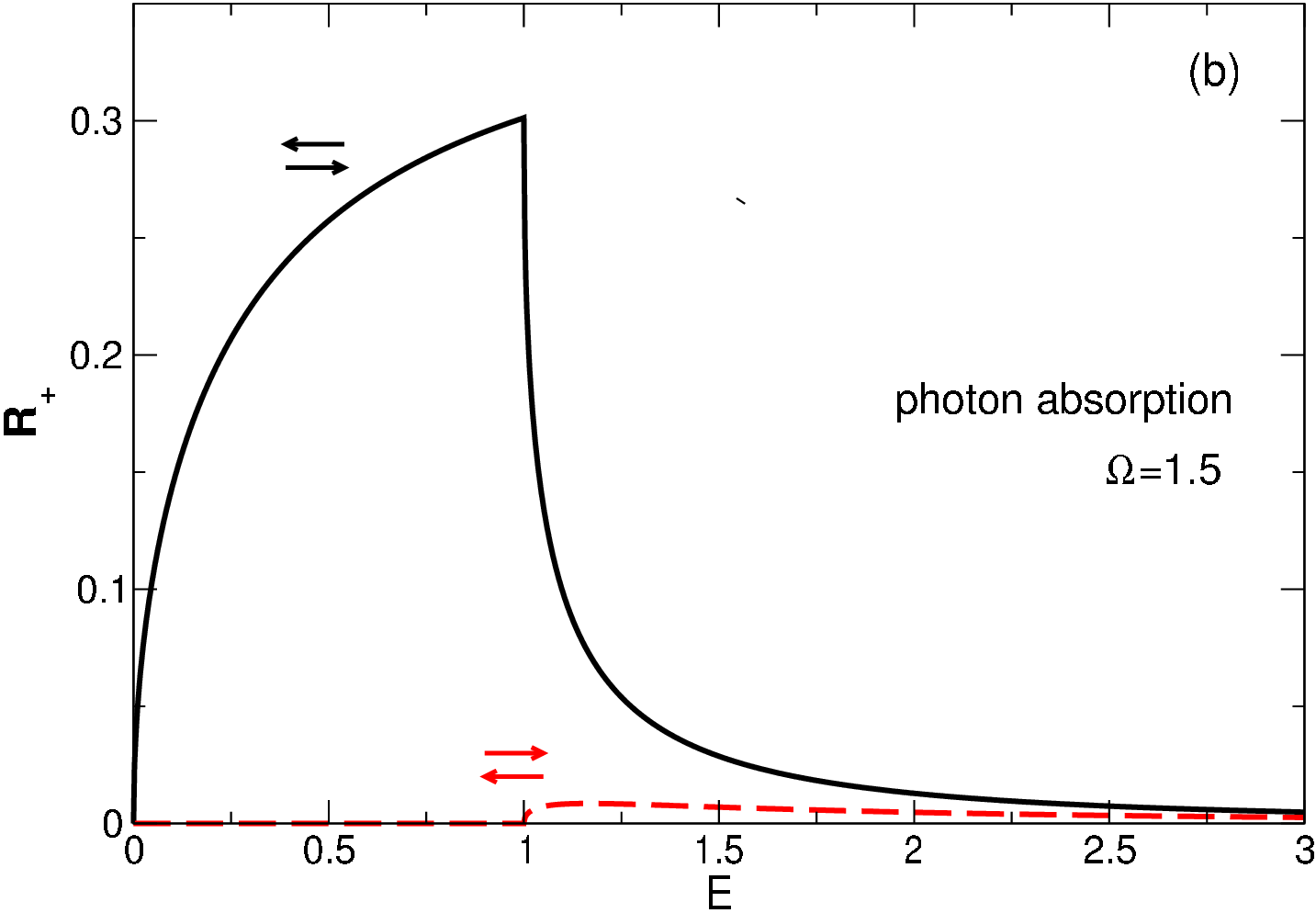}
\caption{\label{fig:R+} The functions $\mathsf{R}_{+}^{\leftrightarrows}({\cal E},\Omega)$ and $\mathsf{R}_{+}^{\rightleftarrows}({\cal E},\Omega)$ defined by equations (\ref{calR-><-}) and (\ref{calR<-->}) for (a) $\Omega<1$ and (b) $\Omega>1$.}
\end{figure}

\begin{figure}
\includegraphics[width=0.49\textwidth]{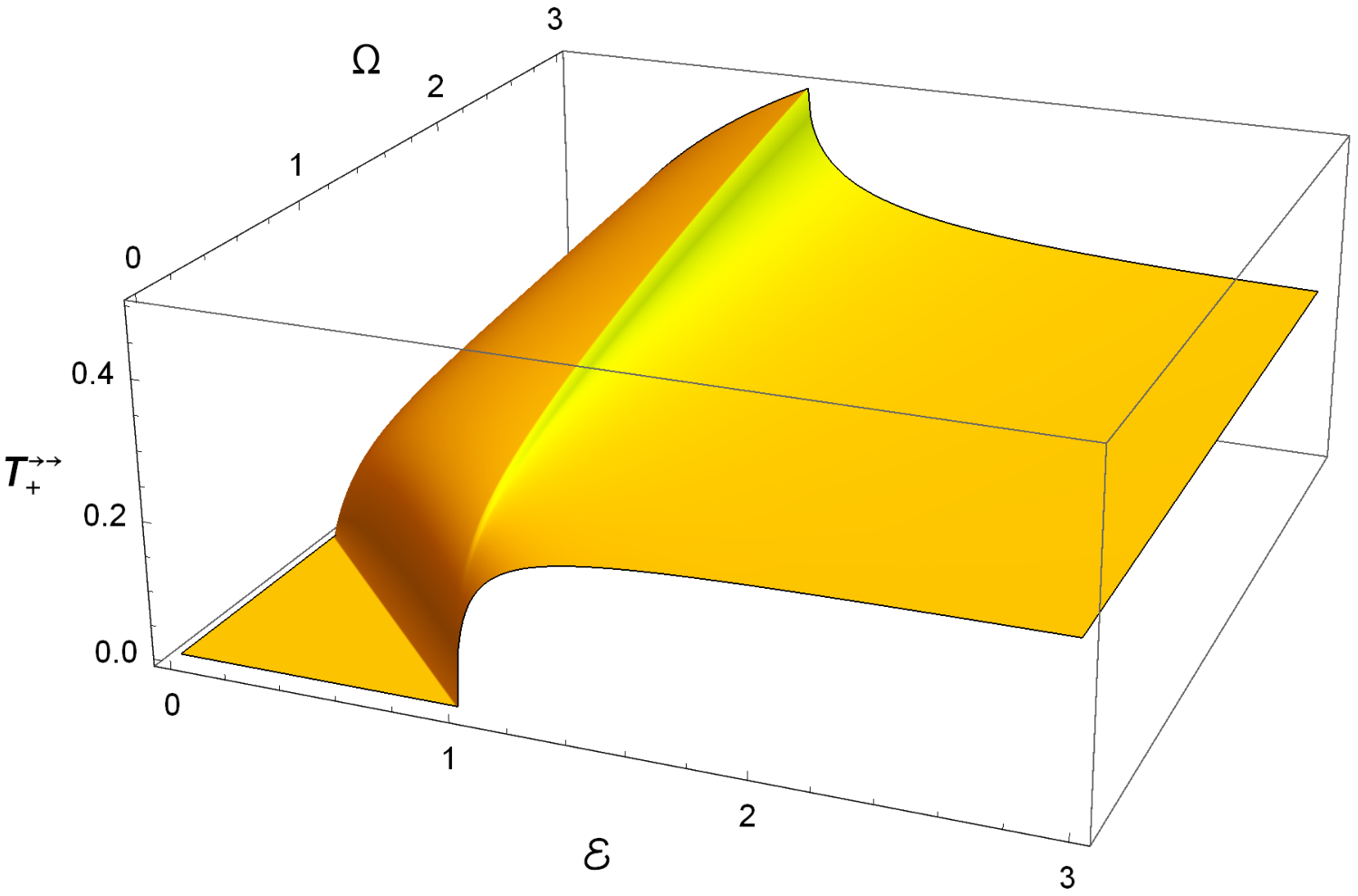}
\includegraphics[width=0.49\textwidth]{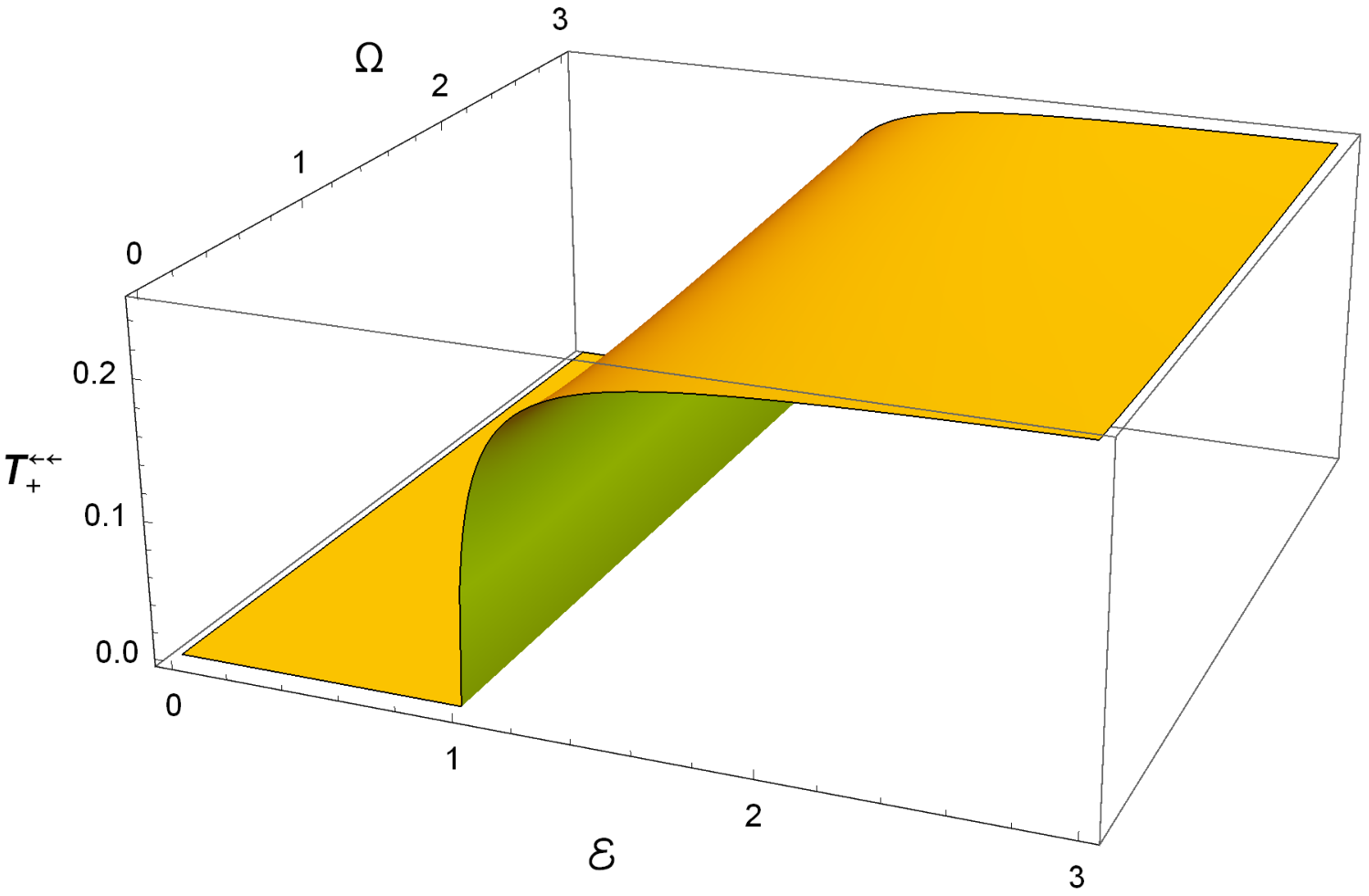}
\includegraphics[width=0.49\textwidth]{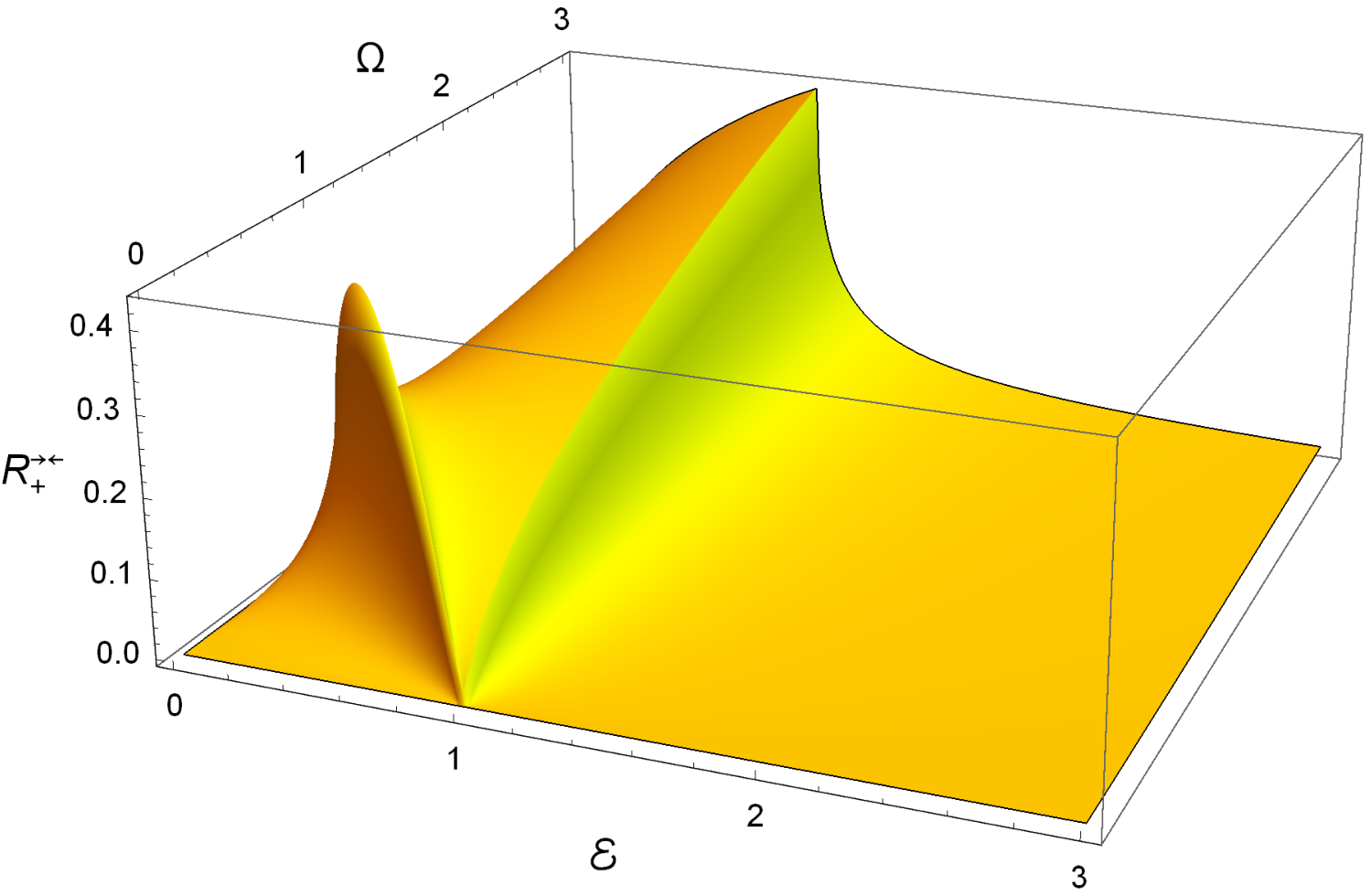}
\includegraphics[width=0.49\textwidth]{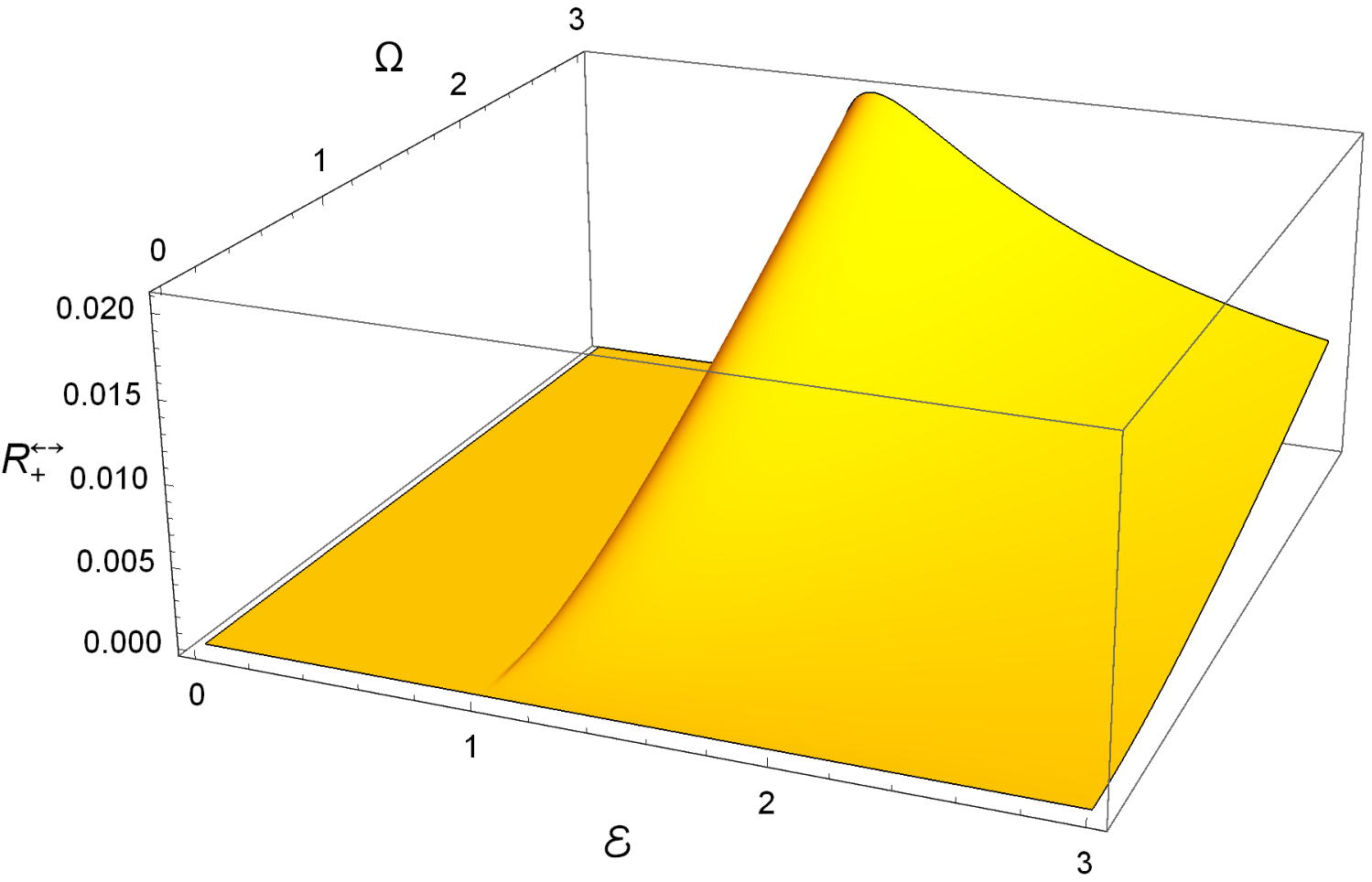}
\caption{\label{fig:TR+_3D} The dimensionless transmission and reflection functions $\mathsf{T}_{+}^{\rightrightarrows}({\cal E},\Omega)$,  $\mathsf{T}_{+}^{\leftleftarrows}({\cal E},\Omega)$, $\mathsf{R}_{+}^{\leftrightarrows}({\cal E},\Omega)$, and $\mathsf{R}_{+}^{\rightleftarrows}({\cal E},\Omega)$, relevant for the absorption of photons, as functions of dimensionless electron (${\cal E}$) and photon ($\Omega$) energies.}
\end{figure}

Figure \ref{fig:R+} illustrates the energy dependence of the reflection functions $\mathsf{R}_{+}^{\leftrightarrows}({\cal E},\Omega)$ and $\mathsf{R}_{+}^{\rightleftarrows}({\cal E},\Omega)$ under the same conditions as in the previous figure. The reflection coefficient for the electron waves running from right to left is always much smaller than that of electrons running from left to right. The function $\mathsf{R}_{+}^{\leftrightarrows}({\cal E},\Omega)$ has a sharp peak at the energy $E=V_B-\hbar\omega$, if $V_B>\hbar\omega$; at energies below and above this value the reflection coefficient quickly decreases. Figure \ref{fig:TR+_3D} shows all four coefficients, $\mathsf{T}_{+}^{\rightrightarrows}({\cal E},\Omega)$, $\mathsf{T}_{+}^{\leftleftarrows}({\cal E},\Omega)$, $\mathsf{R}_{+}^{\leftrightarrows}({\cal E},\Omega)$ and $\mathsf{R}_{+}^{\rightleftarrows}({\cal E},\Omega)$, as 3D plots in dependence on the electron and photon energies ${\cal E}$ and $\Omega$; notice the large difference of the vertical axis scale on the last panel. These plots provide full information about the transmission and reflection coefficients of the photo-excited electrons.

Having found the transmission and reflection coefficients of the photo-excited electrons, we now calculate different physical quantities characterizing the photoresponse of the considered system. 

\subsection{Partial quantum efficiency\label{part-quant-efficiency}}

\begin{figure}
\includegraphics[width=0.49\textwidth]{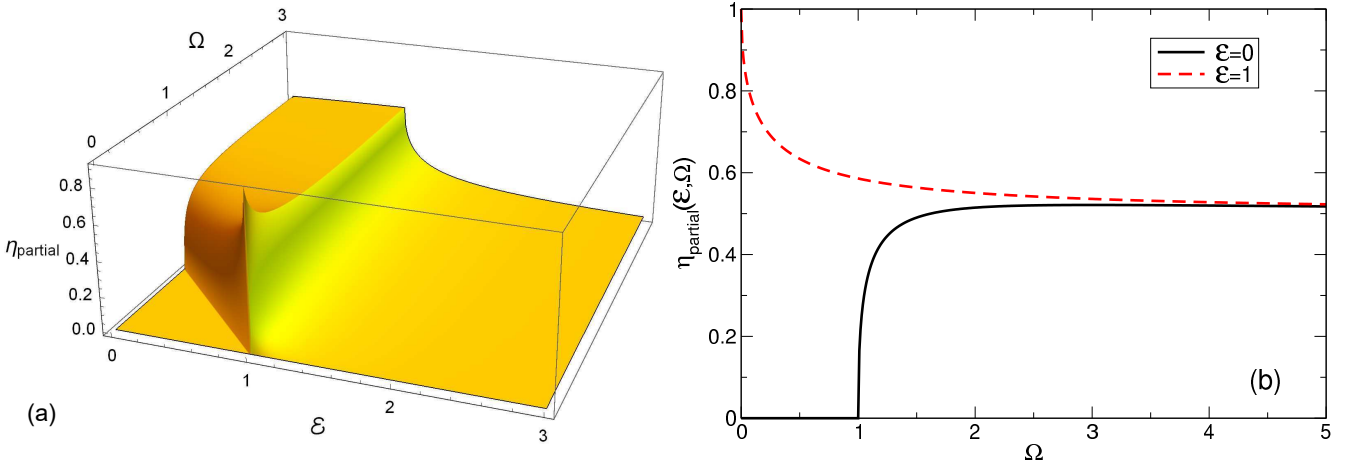}
\includegraphics[width=0.49\textwidth]{QEpartE01.eps}
\caption{\label{fig:QE+} The partial quantum efficiency defined by Eq. (\ref{QEpartial}): (a) a 3D plot as a function of ${\cal E}$ and $\Omega$ and (b) a 2D plot showing $\eta_{\textrm{partial}}$ along the lines ${\cal E}=0$ and ${\cal E}=1$ as a function of $\Omega$.}
\end{figure}

An electron with the energy $E$, moving to the potential step from the left, has the opportunity to absorb a photon and go back to the left, with the probability $R_{E,n}^{+\leftrightarrows}$, or to continue moving to the right, with the probability $T_{E,n}^{+\rightrightarrows}$. In the latter case it contributes to the photocurrent. An electron with the energy $E$, moving to the potential step from the right, has the opportunity to absorb a photon and go back to the right, with the probability $R_{E,n}^{+\rightleftarrows}$, or to continue moving to the left, with the probability $T_{E,n}^{+\leftleftarrows}$. In the latter case it also contributes to the photocurrent but its contribution should be subtracted from that of the right-moving electron. The total probability to absorb a photon for both electrons with the energy $E$ is proportional to the sum $T_{E,n}^{+\rightrightarrows}+ T_{E,n}^{+\leftleftarrows}+ R_{E,n}^{+\leftrightarrows}+ R_{E,n}^{+\rightleftarrows}$. The ratio
\be 
\eta_{\textrm{partial}} ({\cal E},\Omega)=
\frac
{\left|T_{E,n}^{+\rightrightarrows}- T_{E,n}^{+\leftleftarrows}\right|}
{T_{E,n}^{+\rightrightarrows}+ T_{E,n}^{+\leftleftarrows}+ R_{E,n}^{+\leftrightarrows}+ R_{E,n}^{+\rightleftarrows}}
=\frac{\left|\mathsf{T}_{+}^{\rightrightarrows}({\cal E},\Omega)-\mathsf{T}_{+}^{\leftleftarrows}({\cal E},\Omega)\right|}{\mathsf{T}_{+}^{\rightrightarrows}({\cal E},\Omega)+\mathsf{T}_{+}^{\leftleftarrows}({\cal E},\Omega)+\mathsf{R}_{+}^{\leftrightarrows}({\cal E},\Omega)+\mathsf{R}_{+}^{\rightleftarrows}({\cal E},\Omega)}
\label{QEpartial}
\ee
can be considered as a \textit{partial quantum efficiency} of the structure, which refers to a single electron with the energy $E$. Within the first-order perturbation theory this quantity does not depend of the parameter $\alpha$ and illustrates the potential functionality of the in-plane photoelectric effect for detection of electromagnetic radiation. Figure \ref{fig:QE+}(a) shows the quantity (\ref{QEpartial}) as a function of the electron and photon energies, measured in units of the barrier height $V_B$. The quantum efficiency vanishes in the area ${\cal E}+\Omega<1$, where the energy of the photoexcited electron $E+\hbar\omega$ is insufficient to overcome the potential barrier $V_B$. At ${\cal E}+\Omega>1$ and ${\cal E}< 1$ (under-barrier electrons) $\eta_{\textrm{partial}} ({\cal E},\Omega)$ quickly grows and achieves values around $0.5$. At low frequencies, $\Omega\ll 1$, and energies close to the barrier height, ${\cal E}\simeq 1$, the value of $\eta_{\textrm{partial}} ({\cal E},\Omega)$ even exceeds $0.5$ and tends to unity in the limit $\Omega\to 0$ and ${\cal E}\to 1$. Two characteristic lines ${\cal E}=0$ and ${\cal E}=1$ demarcate the area of high $\eta_{\textrm{partial}}$. Along these lines the partial quantum efficiencies $\eta_{\textrm{partial}} (1,\Omega)$ and $\eta_{\textrm{partial}} (0,\Omega)$ are described by the formulas
\be 
\eta_{\textrm{partial}} (1,\Omega)
=\frac{1} {1+\sqrt{\frac{\Omega}{1+\Omega}}},
\label{QEpartial_E1}
\ee
\be 
\eta_{\textrm{partial}} (0,\Omega)=
\frac{ \Theta(\Omega-1)}
{1 +\frac{\Omega}{1+\Omega}
\sqrt{\frac{\Omega}{\Omega-1}}  }.
\label{QEpartial_E0}
\ee
The dependencies (\ref{QEpartial_E1}) and (\ref{QEpartial_E0}) are shown in Figure \ref{fig:QE+}(b). The quantity $\eta_{\textrm{partial}} (0,\Omega)$ has a maximum at $\Omega=3$ which equals $\eta_{\textrm{partial}} (0,3)=\left[1+(3/4)\sqrt{3/2}\right]^{-1}\approx 0.5212$.

The partial quantum efficiency describes the ability of a single electron with the energy $E$ to contribute to the photocurrent. In order to get the full quantum efficiency, the one-electron contributions should be averaged over the equilibrium Fermi distributions of electrons under the left and right gates. This will be done below in Section \ref{sec:quantefficiency}.

\subsection{Photocurrent\label{sec:photocurrent}}

\subsubsection{General formulas\label{sec:PcurrentGeneral}}

Let us now consider the photocurrent response of the system. Assume again, like in Section \ref{sec:quant-conductance}, that different voltages $U_L$ and $U_R$ are applied to the left and right gates, and a dc voltage $U_{sd}$ is applied between the source and drain contacts, Figure \ref{fig:SDcircuit}. In addition, assume that the structure is irradiated by electromagnetic waves polarized along the $x$- and propagating along the $z$-direction. Then the total current, flowing in the system in the $x$-direction, will be given by the sum $I=I^{(0)}+I^{(1)}$, where the zeroth-order contribution (\ref{current0order-general}) has been calculated in Section \ref{sec:quant-conductance}, while the first-order current, similar to (\ref{current0order-general}), is given by
\be 
I^{(1)} =-\frac{e}{\pi\hbar} \sum_{n=1}^\infty \sum_\pm
\int_{-\infty}^\infty  dE 
\left[
T_{En}^{\pm\rightrightarrows} F(E-\mu_S,T)\Big(1-F(E\pm \hbar\omega-\mu_D,T)\Big)- 
T_{En}^{\pm\leftleftarrows} F(E-\mu_D,T)\Big(1-F(E\pm \hbar\omega-\mu_S,T)\Big)
\right].
\label{current1order-general}
\ee
In this formula we still include the terms corresponding to the emission (the superscripts $-$) and absorption (the superscripts $+$) of photons. Equation (\ref{current1order-general}) contains all photoelectric response phenomena in the first-order perturbation theory, including the photocurrent ($\mu_S=\mu_D$) and photoconductivity ($\mu_S\neq \mu_D$) effects. 

In this paper we will focus only on the photocurrent effect postponing the analysis of other photoelectric phenomena to subsequent publications. The photocurrent arises in the system without a source-drain bias, when $\mu_S=\mu_D=\mu_0$. Then the zeroth-order current $I^{(0)}$ vanishes and we get $I=I^{(1)}$, where
\be 
I^{(1)} =-\frac{e}{\pi\hbar} 
\sum_{n=1}^\infty \sum_\pm
\int_{-\infty}^\infty  dE 
\Big(T_{En}^{\pm\rightrightarrows} - T_{En}^{\pm\leftleftarrows} \Big)
F(E-\mu_0,T)\Big(1-F(E\pm \hbar\omega-\mu_0,T)\Big).
\label{photocurrent}
\ee
Substituting the transmission coefficients (\ref{T->->}) and (\ref{T<-<-}) into the formula (\ref{photocurrent}) we get
\ba 
I^{(1)} &=&-\frac{e}{\pi\hbar} \left(\frac {e\Delta\Phi_{ac}}{\hbar\omega}\right)^2
\sum_{n=1}^\infty \sum_\pm
\int_{-\infty}^\infty  dE 
F(E-\mu_0,T)\Big(1-F(E\pm \hbar\omega-\mu_0,T)\Big)
\nonumber \\ &\times&
\left[
\mathsf{T}_{\pm}^{\rightrightarrows}\left(\frac{E-E_Wn^2-V_L}{V_B},\frac{\hbar\omega}{V_B}\right) - \mathsf{T}_{\pm}^{\leftleftarrows}\left(\frac{E-E_Wn^2-V_L}{V_B},\frac{\hbar\omega}{V_B}\right)
\right],
\label{photocurrent-1}
\ea
where the functions $\mathsf{T}_{\pm}^{\rightrightarrows}$ and $\mathsf{T}_{\pm}^{\leftleftarrows}$ are determined by Eqs. (\ref{calT->->}) and (\ref{calT<-<-}). Equation (\ref{photocurrent-1}) gives a general analytical expression for the photocurrent $I^{(1)}$ in the considered structure.  

\subsubsection{Special case: A macroscopically wide 2D channel at zero temperature\label{sec:PcurrentSpecial}}

As seen from the formula (\ref{photocurrent-1}), the photocurrent $I^{(1)}$ depends on several energy scales: (a) the equilibrium chemical potential $\mu_0$, (b) parameters of the potential step ($V_L$, $V_B$), (c) the photon energy $\hbar\omega$, (d) the thermal energy $T$, and (e) the transverse quantization energy $E_W$. In a typical GaAs/AlGaAs heterostructure with a 2DEG the energy $\mu_0$ is on the order of 10 to 40 meV and the potential step parameters can be experimentally tuned in the same range (tens of meV). The photon energy at THz frequencies is of the same order: the frequencies of $f\sim 1-3$ THz correspond to $\hbar\omega\sim 4-12$ meV. Two other energy parameters, $T$ and $E_W$, are significantly smaller under the conditions of the experiment \cite{Michailow22}: the thermal energy at $T= 9$ K (the temperature in \cite{Michailow22}) is less than 0.8 meV, and the quantization energy in a micron wide 2D channel is on the $\mu$eV scale ($E_W\approx 5.6$ $\mu$eV at $W\approx 1$ $\mu$m). The latter fact means that $E_W\ll \mu_0$, i.e., the number $N_{\rm 1D}$ of occupied quasi-1D electron subbands in the 2D channel is much larger than 1.

In order to clarify the physics of the IPPE effect we will consider in this paper only the special case, when two of the energy parameters, the temperature $T$ and the transverse quantization energy $E_W$, tend to zero. More general situations will be investigated later.

First, consider the limit $E_W\to 0$. Then the sum over $n$ in Eq. (\ref{photocurrent-1}) can be replaced by the integral, according to the rule $\sum_{n=1}^\infty f(n^2)\approx \int_0^\infty d\nu f(\nu^2)$. Introducing the variables $E_y=\hbar^2 \pi^2 \nu^2/2mW^2$, $E_x=E-E_y$, and changing the order of integration, we obtain
\ba 
I^{(1)} &\approx&
-\frac{e}{2\pi\hbar} \left(\frac {e\Delta\Phi_{ac}}{\hbar\omega}\right)^2
\sum_\pm \frac{1}{\sqrt{E_W}}
\int_{-\infty}^\infty  dE_x \left[
\mathsf{T}_{\pm}^{\rightrightarrows}\left(\frac{E_x-V_L}{V_B},\frac{\hbar\omega}{V_B}\right) - \mathsf{T}_{\pm}^{\leftleftarrows}\left(\frac{E_x-V_L}{V_B},\frac{\hbar\omega}{V_B}\right)
\right]
\nonumber \\ &\times&
\int_{0}^\infty \frac{dE_y}{\sqrt{E_y}}  F(E_x+E_y-\mu_0,T)\Big(1-F(E_x+E_y\pm \hbar\omega-\mu_0,T)\Big).
\label{photocurrent-widechannel}
\ea
The integral over $dE_y$ in (\ref{photocurrent-widechannel}) can now be easily calculated at $T\to 0$. We see that in this case the emission term with the sign ``$-$'' gives zero contribution, due to the Fermi-function dependent factors. The term describing the photon absorption gives
\be
\int_{0}^\infty \frac{dE_y}{\sqrt{E_y}}  F(E_x+E_y-\mu_0,T)\Big(1-F(E_x+E_y+ \hbar\omega-\mu_0,T)\Big)=2\Theta(X)\Big(\sqrt{X}-\sqrt{\max\{0,X-\hbar\omega\}}\Big),
\ee
where $X=\mu_0-E_x$. Further, introducing the local chemical potentials under the left and right gates $\mu_{L,R}=\mu_0-V_{L,R}$, we get after some algebra \ba 
I^{(1)} =-
\frac{e\omega }{\pi}\left(\frac {e\Delta\Phi_{ac}}{\hbar\omega}\right)^2
\sqrt{\frac{\hbar\omega}{E_W}}
{\cal J}\left(\frac{\mu_{L}}{\hbar\omega},\frac{\mu_{R}}{\hbar\omega}\right),
\label{photocurrent-wideT0}
\ea
where the dimensionless function ${\cal J}$ is defined as 
\be 
{\cal J}(\zeta_{L},\zeta_{R})= 
\int_{0}^\infty  dx  \left(\sqrt{x}-\sqrt{\max\{0,x-1\}}\right) 
\Big[A(x,\zeta_L,\zeta_R)-A(x,\zeta_R,\zeta_L)\Big],
\label{funcalJ}
\ee
$\zeta_{L,R}=\mu_{L,R}/\hbar\omega$, and 
\be 
A(x,\zeta_L,\zeta_R)=\Theta(\zeta_L- x)\Theta(\zeta_R+1- x) \frac{\sqrt{\zeta_L- x} \sqrt{\zeta_R+1- x}
\left|\sqrt{\zeta_R- x}+\sqrt{\zeta_L+1- x}\right|^2}
{\left|\sqrt{\zeta_L- x}+\sqrt{\zeta_R- x} \right|^2 \left|\sqrt{\zeta_L+1- x} +\sqrt{\zeta_R+1- x}\right|^2}.\label{funA}
\ee

The formula (\ref{photocurrent-wideT0}) gives the IPPE photocurrent, generated at $T=0$ in a macroscopically wide ($E_W\to 0$) sample, under irradiation by the electromagnetic waves with the frequency $\omega$. The photocurrent (\ref{photocurrent-wideT0}) is a product of several factors. The term $e\omega/\pi=2ef$ has the dimension of current and equals $0.32$ $\mu$A at the frequency of 1 THz. The factor $\sqrt{\hbar\omega/E_W}\propto W$ determines the photocurrent dependence on the width of the 2D layer, $I^{(1)} \propto W$. If the 2D channel width is of a micron scale, this factor is much larger than unity at THz frequencies: for example, if $W=1$ $\mu$m and $f=1$ THz, then $\hbar\omega=4.12$ meV, $E_W=5.58$ $\mu$eV, and $\sqrt{\hbar\omega/E_W}\approx 27.2$ in a GaAs/AlGaAs quantum well. The factor $\alpha=\left(e\Delta\Phi_{ac}/\hbar\omega\right)^2$ is the perturbation theory parameter as was discussed in Section \ref{sec:TR+-}. 

\begin{figure}[ht]
\includegraphics[width=0.49\textwidth]{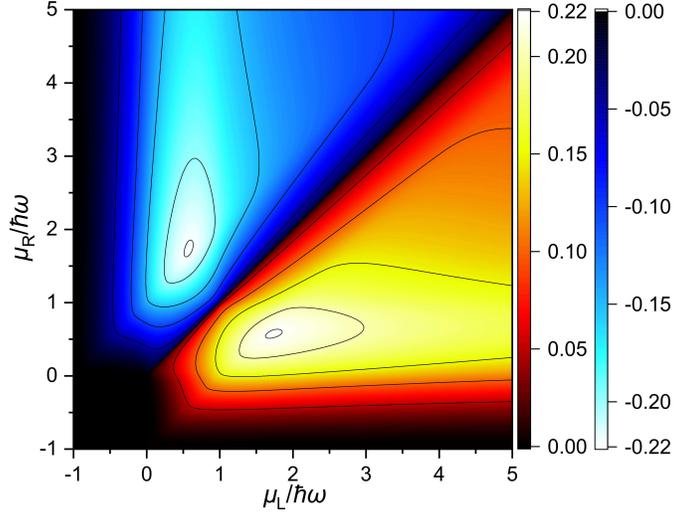} 
\caption{\label{fig:IphT=0} The function ${\cal J}$, Eq. (\ref{photocurrent-wideT0}), which determines the photocurrent response of the system at $T=0$ in the quasiclassical limit $E_W\to 0$, in dependence on $\mu_L/\hbar\omega$ and $\mu_R/\hbar\omega$. The maxima of the function $|{\cal J}|$ are located at the points $(\zeta_L,\zeta_R)=(1.725,0.575)$ and $(\zeta_L,\zeta_R)=(0.575,1.725)$, see Eq. (\ref{optimalMuLR}).}
\end{figure}

The universal function ${\cal J}$ depends on two dimensionless arguments: the chemical potentials of electrons under the left and right gates $\mu_L$ and $\mu_R$, normalized to the photon energy $\hbar\omega$. It is shown in Figure \ref{fig:IphT=0}. One sees that it vanishes at the diagonal $\mu_L=\mu_R$, changes sign when $\mu_L$ and $\mu_R$ are interchanged, and becomes positive at $\mu_L>\mu_R$ meaning that the photoexcited electrons flow onto the potential step, from the area with a larger electron density to the area with a smaller electron density. This function equals zero in the left bottom corner of the $\mu_L$-$\mu_R$ plane where both chemical potentials are negative. In the areas where only one of the chemical potentials is negative ($\mu_L<0<\mu_R$ or $\mu_R<0<\mu_L$), i.e., in the pinch-off regime, the photocurrent decreases and finally vanishes when $\zeta_L$ or $\zeta_R$ become smaller than $-1$. The general shape of the photocurrent dependence on the chemical potentials, Figure \ref{fig:IphT=0}, is in good qualitative agreement with the experimental observations in Ref. \cite{Michailow22}. The numerical data for the function ${\cal J}(\zeta_L,\zeta_R)$ (with negative sign) can be found in Ref. \cite{Michailow22-data}.

\begin{figure}[ht]
\includegraphics[width=0.48\textwidth]{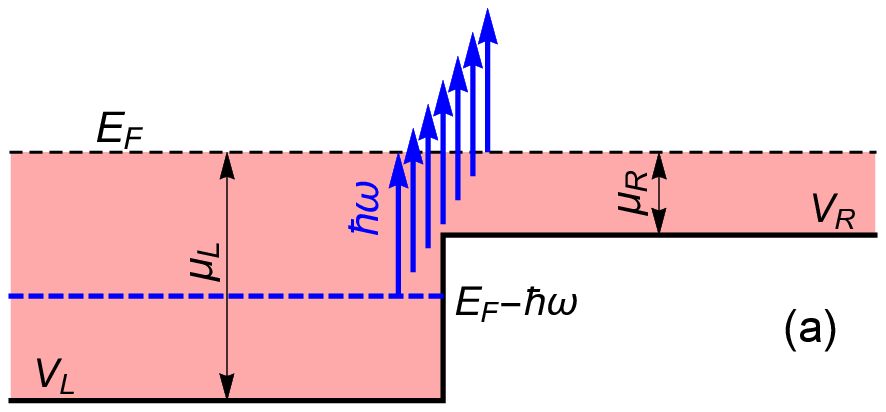} \hfill
\includegraphics[width=0.48\textwidth]{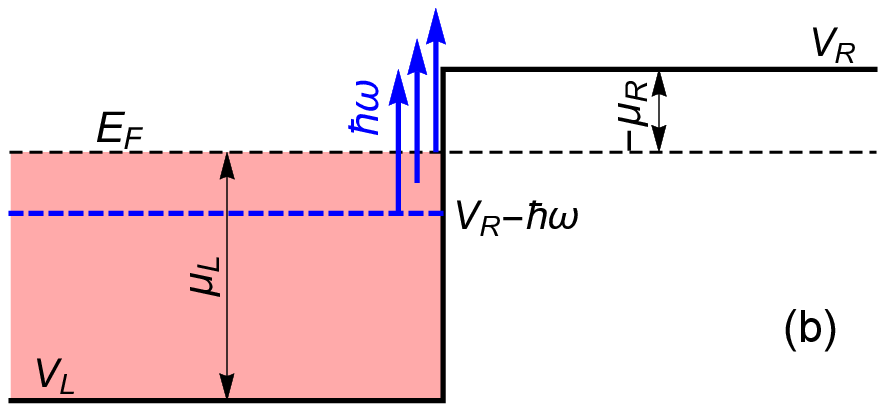} 
\caption{\label{fig:vertical_transitions_atT=0} Vertical transitions leading to the photocurrent (a) in the point of maximum photocurrent $\mu_L/\hbar\omega\approx 1.725$, $\mu_R/\hbar\omega\approx 0.575$, and (b) in the depletion region $\mu_L/\hbar\omega\approx 1.725$, $\mu_R/\hbar\omega\approx -0.575$. Pink areas: occupied states; white areas above the Fermi energy and $V_R$: empty states. Blue arrows (horizontally shifted for clarity) symbolize the absorption of THz photons. The horizontal dashed blue line shows the lowest electron energy from which the transitions are possible: (a) $E_F-\hbar\omega$, if $E_F>V_R$, and (b) $V_R-\hbar\omega$, if $E_F<V_R$. The range of initial energies of electrons that are able to absorb a photon and contribute to the photocurrent equals (a) $\hbar\omega$ at $E_F>V_R$, and (b) $\hbar\omega-(V_R-E_F)$, at $E_F<V_R$; it is broader in the case (a).}
\end{figure}

In contrast to the conventional photoelectric effect, in which one of the chemical potentials is always negative, Figure \ref{fig:vertical_transitions_atT=0}(b), the IPPE photocurrent has a maximum when \textit{both} chemical potentials are \textit{positive} and the 2D gas is degenerate in both parts of the device, left and right from the potential step. The maximum of the photocurrent function $|{\cal J}|$ is thus achieved at a \textit{negative} ``work function'', as was discussed in the Introduction, in the point 
\be 
\zeta_L=\frac{\mu_L}{\hbar\omega}\approx 1.725, \ \ \ \zeta_R=\frac{\mu_R}{\hbar\omega}\approx 0.575
\label{optimalMuLR}
\ee 
(or vice versa), Figure \ref{fig:IphT=0}. The function $|{\cal J}|=|{\cal J}_{\max}|$ in this point approximately equals $0.22079$. This maximum is very broad: the region of the $\zeta_L$-$\zeta_R$ plane where $|{\cal J}|$ is larger than 0.2, i.e., smaller than $|{\cal J}_{\max}|$ by only 10\%, covers a wide area in which the normalized chemical potentials $\zeta_L$ and $\zeta_R$ deviate from the optimal point (\ref{optimalMuLR}) by $40-50$\%. 

Figure \ref{fig:vertical_transitions_atT=0} helps to understand the physical reason of this remarkable feature. It illustrates the generation of the photocurrent due to the vertical electronic transitions near the potential step (a) in the point of the photocurrent maximum $\mu_L/\hbar\omega\approx 1.725$, $\mu_R/\hbar\omega\approx 0.575$, and (b) in the conventional case when one of the chemical potentials is negative, e.g. at $\mu_L/\hbar\omega\approx 1.725$, $\mu_R/\hbar\omega\approx -0.575$. The horizontal dashed blue line shows the lowest electron energy from which the transitions are possible. In the case (a), when both chemical potentials are positive, more energy states contribute to the photocurrent, as compared to the case (b), where the number of electrons which are able to overcome the potential barrier is substantially smaller. Figure \ref{fig:vertical_transitions_atT=0}(a) also helps to understand why the photocurrent maximum is so broad: one sees that moderate changes of the chemical potentials around the optimal point do not significantly change the range of energy states which may contribute to the photocurrent. It is also qualitatively clear that, if the Fermi distribution edge is blurred by a finite temperature, this should not lead to big changes of the photoresponse.

\subsubsection{Equivalent circuit of the PETS detector \label{sec:equivcircuit}}

\begin{figure}[ht]
\includegraphics[width=0.33\textwidth]{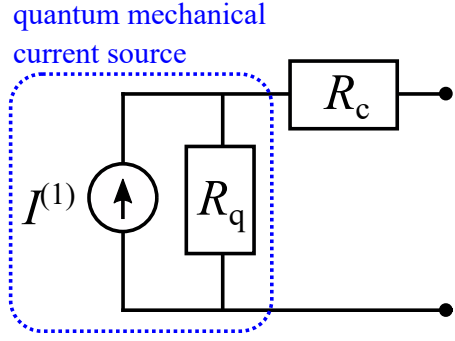}
\caption{\label{fig:equivcirc} The equivalent circuit of the PETS detector based on the IPPE effect. $I^{(1)}$ is the ideal current source given by Eq. (\ref{photocurrent-wideT0}). $R_q=1/\sigma_q$ is the internal (quantum) resistance related to the conductance $\sigma_q$ determined by Eq. (\ref{quant-conduct}). $R_c$ is the classical resistance describing the 2DEG areas connecting the quantum region of the potential step with the source and drain contacts, as well as any classical resistances of the connecting leads, such as, e.g., Ohmic contacts.
} 
\end{figure}

Now we can discuss the equivalent circuit of the PETS detector. As seen from our results, the photocurrent has a quantum-mechanical origin and arises in the microscopic area (of the width $|x|\lesssim b\ll l_{\mathrm{mfp}}$) at the potential step $V_0(x)$. The PETS detector based on the IPPE effect can therefore be described by the equivalent circuit shown in Figure \ref{fig:equivcirc}. The short microscopic area that generates the photocurrent is presented here by an ideal current source $I^{(1)}$, Eq. (\ref{photocurrent-wideT0}), connected in parallel with the internal (quantum) resistance $R_q=1/\sigma_q$, with the quantum conductance $\sigma_q$ calculated in Eq. (\ref{quant-conduct}). $R_c$ consists of the classical resistance of the 2DEG calculated within the Drude model, as well as any classical resistances of the connecting leads (e.g. Ohmic contacts). This resistance is device-specific, and the estimation of the experimentally measured photocurrent requires consideration of the loading of the current source depicted in Fig. \ref{fig:equivcirc} with the experimental load resistance.

\subsection{Quantum efficiency\label{sec:quantefficiency}}

\begin{figure}[ht]
\includegraphics[width=0.49\textwidth]{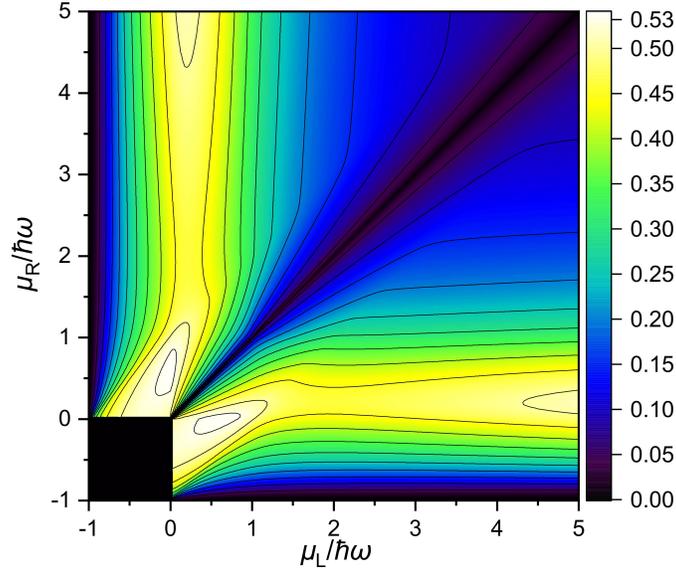} 
\caption{\label{fig:QE-T0Ew0} The quantum efficiency (\ref{QE-T0}) as a function of $\mu_L/\hbar\omega$ and $\mu_R/\hbar\omega$ at $T=0$ and $E_W\to 0$.
} 
\end{figure}

In Section \ref{part-quant-efficiency} we have defined the \textit{partial} quantum efficiency $\eta_{\rm partial}({\cal E},\Omega)$, the quantity which determines how efficiently an electron with the energy $E$ may contribute to the photocurrent after absorption of a THz photon. In order to get the total quantum efficiency $\eta(\zeta_L,\zeta_R)$ which refers to the ensemble of particles, we have to average all terms in the definition of $\eta_{\rm partial}$, both in the nominator and in the denominator of (\ref{QEpartial}), over the Fermi distributions. Repeating the same algebraic transformations as in Section \ref{sec:PcurrentSpecial} we get at $E_W\to 0$ and $T=0$
\be 
\eta (\zeta_L,\zeta_R)=\frac{\left|\int_{0}^\infty  dx \left(\sqrt{x}-\sqrt{\max\{0,x-1\}}\right)
\Big[A(x,\zeta_L,\zeta_R)-A(x,\zeta_R,\zeta_L)\Big]\right|}
{\int_{0}^\infty  dx \left(\sqrt{x}-\sqrt{\max\{0,x-1\}}\right)
\Big[A(x,\zeta_L,\zeta_R)+A(x,\zeta_R,\zeta_L)+B(x,\zeta_L,\zeta_R)+B(x,\zeta_R,\zeta_L)\Big]}
\label{QE-T0}
\ee
where the function $B(x,\zeta_L,\zeta_R)$ is defined as follows
\be 
B(x,\zeta_L,\zeta_R)=
\Theta(\zeta_L-x)\Theta(\zeta_L+1-x)
\frac{\sqrt{\zeta_L-x}\sqrt{\zeta_L+1-x}
\left|\sqrt{\zeta_R-x}-\sqrt{\zeta_R+1-x}\right|^2}
{\left|\sqrt{\zeta_R-x}+\sqrt{\zeta_L-x} \right|^2 
\left|\sqrt{\zeta_L+1-x} +\sqrt{\zeta_R+1-x}\right|^2}.
\ee 
The quantum efficiency $\eta(\zeta_L,\zeta_R)$ is symmetric, $\eta(\zeta_L,\zeta_R)=\eta(\zeta_R,\zeta_L)$.

Figure \ref{fig:QE-T0Ew0} shows the quantum efficiency $\eta(\zeta_L,\zeta_R)$ as a function of the left and right chemical potentials normalized to the photon energy $\hbar\omega$. In the area where both chemical potentials are negative, the function $\eta(\zeta_L,\zeta_R)$ vanishes. The maximum efficiency of the photon -- electron transformation is achieved at $\zeta_L\approx -0.175$ and $\zeta_R\approx 0.375$ and is about 53\%. Notice that the $(\zeta_L,\zeta_R)$-area where the quantum efficiency is maximal does not coincide with the $(\zeta_L,\zeta_R)$-area where the photocurrent is maximal. In particular, in the point of the photocurrent maximum, $(\zeta_L,\zeta_R)=(1.725,0.575)$, the quantum efficiency is about 42.3\%. 

\subsection{Internal responsivity and frequency dependence of the photoresponse \label{sec:responsivity}}

The formula (\ref{photocurrent-wideT0}) for the photocurrent can be rewritten in the following useful form:
\ba 
\frac{I^{(1)}}{W(\Delta\Phi_{ac})^2} =
-
\frac{e }{\hbar}{\cal R}\left(\frac{\mu_{L}}{\hbar\omega},\frac{\mu_{R}}{\hbar\omega},\frac{G}{\hbar\omega}\right),
\label{photocurrent-wideT0-inR}
\ea
where we have defined one more quantity having the unit of energy,
\be 
G=\frac{2me^4}{\pi^4\hbar^2},
\ee
and introduced a function 
\be 
{\cal R}\left(\frac{\mu_{L}}{\hbar\omega},\frac{\mu_{R}}{\hbar\omega},\frac{G}{\hbar\omega}\right)= \sqrt{\frac{G}{\hbar\omega}}
{\cal J}\left(\frac{\mu_{L}}{\hbar\omega},\frac{\mu_{R}}{\hbar\omega}\right).
\label{int-responsivity}
\ee 
The energy $G$ coincides, up to a numerical constant, with the effective Rydberg energy in a semiconductor. It depends only on one material parameter (the electron effective mass $m$) and equals $G=37.43$ meV in GaAs ($m_{\rm GaAs}=0.067m_0$). The dimensionless function ${\cal R}$ can be treated as an \textit{internal responsivity} of the IPPE effect: it determines the current density $I^{(1)}/W$, which is independent of the channel width $W$, and is generated in the 2D channel by a given squared ac potential difference $(\Delta\Phi_{ac})^2$ in the gap between the antenna wings. Its dependence on the chemical potentials is the same as that of the photocurrent ${\cal J}$, see Figure \ref{fig:IphT=0}. The prefactor $e/\hbar$ in (\ref{photocurrent-wideT0-inR}) depends only on fundamental constants and numerically equals
\be 
\frac{e}{\hbar}\approx 1.7\frac {\textrm{kA}}{\textrm{cm V}^2}. \label{e/h-numeric}
\ee 

What is the frequency dependence of the internal responsivity ${\cal R}$? What is the fundamental maximum value of the function ${\cal R}$? These questions can be answered as follows. The frequency $\omega$ enters all three dimensionless arguments of the function ${\cal R}$, $\mu_{L}/\hbar\omega$, $\mu_{R}/\hbar\omega$, and $G/\hbar\omega$. However, aiming for the maximum photoresponse of the system, we can fit, at any given frequency, the left and right gate voltages to the points $\mu_L^{\max}=1.725\hbar\omega$ and $\mu_R^{\max}=0.575\hbar\omega$ (or vice versa), Eq. (\ref{optimalMuLR}), to get the maximum value of the function $|{\cal J}|=|{\cal J}_{\max}|=0.22079$. Then the frequency dependence of ${\cal R}_{\max}$ remains only in the third argument $G/\hbar\omega$ and we get for the maximum internal responsivity of the IPPE effect
\be 
{\cal R}_{\max}\approx \sqrt{\frac{G}{\hbar\omega}}
{\cal J}_{\max}=0.22079\sqrt{\frac{G}{\hbar\omega}}.
\label{max-responsivity}
\ee

\begin{figure}
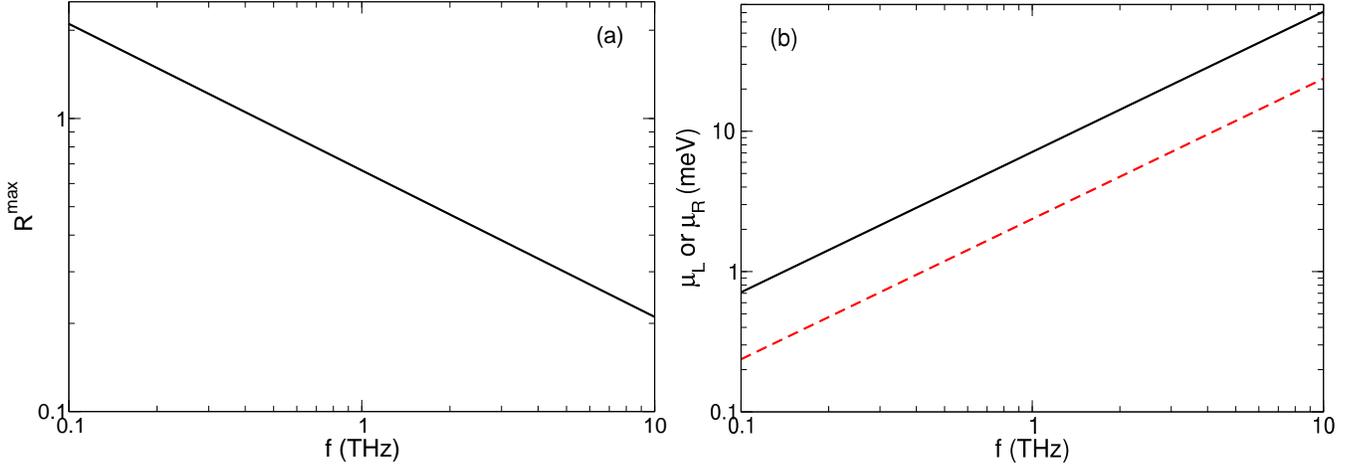

\includegraphics[width=0.49\textwidth]{respf.eps}
\includegraphics[width=0.49\textwidth]{muLRf.eps}
\caption{\label{fig:resp} (a) The maximal value of the function ${\cal R}^{\max}$, defined in Eq. (\ref{max-responsivity}), and (b) the chemical potentials $\mu_L^{\max}$ and $\mu_R^{\max}$, corresponding to the maximal photoresponse, as functions of the frequency $f$ in the interval $0.1-10$ THz.}
\end{figure}

Figure \ref{fig:resp} shows the maximal value of the ${\cal R}$-function, Eq. (\ref{max-responsivity}), as well as the optimal chemical potentials (\ref{optimalMuLR}), as functions of the frequency in the range $0.1-10$ THz. ${\cal R}^{\max}$ slowly falls down with the frequency, ${\cal R}\propto 1/\sqrt{\omega}$, from a value of $\sim 2$ at $f=0.1$ THz down to $\sim 0.2$ at $f=10$ THz. Together with the factor (\ref{e/h-numeric}) this gives the maximal internal responsivity of the IPPE effect in the range of $\sim 3.4$ kA/cmV$^2$ at $f=0.1$ THz and $\sim 0.34$ kA/cmV$^2$ at $f=10$ THz. 

The slow ($\omega^{-1/2}$) frequency dependence of the quantum (in-plane photoelectric) detection mechanism is in contrast to the much stronger frequency drop of single-particle classical mechanisms ($\omega^{-2}$) \cite{Levin15,McColl77}. The IPPE effect is therefore well suited for the detection of radiation in the entire THz frequency range.

\subsection{External responsivity\label{sec:ext-responsivity}}

Equation (\ref{photocurrent-wideT0-inR}) relates the generated photocurrent to the electric potential difference $\Delta\Phi_{ac}\simeq E_{ac}b$ acting on electrons in the gap between the two gates. The ac electric field inside the gap $E_{ac}$ is related, in its turn, to the electric field of the incident electromagnetic wave $E_0$, and hence to its intensity ${\cal I}_0=(c/4\pi)E_0^2$. The ratio $K=E_{ac}/E_0$ of the field amplitudes depends on the antenna design and is a technical question which we do not discuss in this paper. Combining (\ref{photocurrent-wideT0-inR}) and (\ref{max-responsivity}) the current density generated by the incident radiation with the intensity ${\cal I}_0$ can then be written as
\be 
\left|\frac{I^{(1)}/W}{{\cal I}_0}\right|_{\max}  \simeq  2.76(Kb)^2
\frac{e }{\hbar c}\sqrt{\frac{G}{\hbar\omega}}\approx 4.22 K^2\frac{(b[\mu{\rm m}])^2}{\sqrt{f[{\rm THz}]}}\ \frac{\rm mA/cm}{{\rm W/cm}^2}.
\label{max-response}
\ee
The quantity (\ref{max-response}) can be considered as the maximum \textit{external} responsivity, normalized to the incident power density. It determines the photocurrent $I^{(1)}$ of the ideal current source shown on the equivalent circuit in Figure \ref{fig:equivcirc}. 
 
\subsection{Approximations used and their applicability \label{sec:finremark}} 

Let us now discuss how restrictive the approximations made in this paper are and what can be expected beyond their applicability. Our results have been obtained under the conditions $E_W\ll T$, with $T$ being much smaller than all other energy parameters of the problem ($\mu_0$, $\hbar\omega$, etc.). If these conditions are not satisfied, calculations can be performed in a similar way, using the general formulas obtained in Section \ref{sec:PcurrentGeneral} above. In narrow samples one can expect one-dimensional quantization \cite{Wees88,Wharam88} to have an influence on the photocurrent response. With growing temperature the photoresponse will decrease, but due to the reasons mentioned in Section \ref{sec:PcurrentSpecial}, the photocurrent reduction should not be big, at least up to temperatures $T\lesssim\hbar\omega$.

Further, we have assumed that the gap between the gates is much smaller than the mean free path of 2D electrons, $b\ll l_{\rm mfp}$. This condition allows to ignore the scattering of photoexcited electrons in the gap between the gates and consider the photoexcitation process purely quantum-mechanically. If the condition $b\ll l_{\rm mfp}$ starts to be violated, e.g., due to mobility degradation with increasing temperature, then the classical mechanisms which rely on the electron scattering, e.g. the photothermoelectric effect, may become relevant. In general, the interplay of the quantum (IPPE) and classical detection mechanisms (photothermoelectric\cite{Gabor11,Viti16,Castilla19,Viti20}, plasmonic mixing \cite{Dyakonov96,Viti15,Bandurin18,Sun12a}) may be an interesting topic for further theoretical research. 

Finally, we have explored the model of a step-like potential profile seen by electrons on their way from the source to drain. In reality, the potential step will not be step-like, but smooth. The smoothness effects, however, should not strongly influence our results as long as the condition $b\ll l_{\rm mfp}$ remains true. Indeed, without irradiation electrons with the energies $E<V_B$ are reflected from both the step-like and smooth barrier, while at $E>V_B$ they pass through the potential step with the probability close to unity. The only difference between the step-like and the smooth barrier is that the function $\mathsf{T}_{0}({\cal E})$ shown in Figure \ref{fig:TR0} will be steeper in the case of a smooth barrier, see Ref. \cite{Landau3}, problem 3 to \S 25. Similarly, one can expect that energy dependencies of the coefficients $\mathsf{T}_{+}^\rightrightarrows({\cal E},\Omega)$ and $\mathsf{T}_{+}^\leftleftarrows({\cal E},\Omega)$ (Figure \ref{fig:T+}) will also be steeper for a smooth barrier. But this should not substantially influence the photocurrent, as long as the photoexcited electrons have the opportunity to quickly leave the photoexcitation area $|x|\lesssim b/2$, i.e., if their mean free path exceeds the width $b$ between the antenna wings, $b\ll l_{\rm mfp}$.

\section{Summary\label{sec:conclusion}} 

We have presented an analytical theory of the in-plane photoelectric effect in a semiconductor heterostructure with a 2D electron gas. Having solved the time-dependent Schr\"odinger equation for electrons moving in the 2D electron gas in the PETS detector, we have calculated the transmission and reflection coefficients of the photoexcited electrons, the photocurrent, as well as the quantum efficiency, the internal responsivity, and the internal quantum conductance of the device. We have found that the quantum efficiency can be as large as $\simeq 50$\%. We have also shown that the internal responsivity of the PETS detector weakly depends on the frequency, so that it can be used for efficient detection of THz radiation across the entire THz range $\sim 0.1 -10$ THz. 

The theory developed in this work is valid for all structures with 2D electron gases, in which electrons have a parabolic energy dispersion. This includes both traditional semiconductor heterostructures based on III-V compounds or Si/SiO$_2$ MOSFETs, as well as novel, atomically thin semiconductors such as transition metal dichalcogenides, phosphorene, and other similar materials.

\begin{acknowledgments}
S.A.M. acknowledges funding from the European Union's Horizon 2020 research and innovation program Graphene Core 3 under Grant Agreement No. 881603. W.M. thanks Trinity College Cambridge for a Junior Research Fellowship and the Schiff Foundation of the University of Cambridge for a George and Lillian Schiff Studentship. The work at the University of Cambridge has been also funded by EPSRC under the HyperTerahertz grant No. EP/P021859/1.
\end{acknowledgments}

%

\bibliography{}

\begin{thebibliography}{32}%
\makeatletter
\providecommand \@ifxundefined [1]{%
 \@ifx{#1\undefined}
}%
\providecommand \@ifnum [1]{%
 \ifnum #1\expandafter \@firstoftwo
 \else \expandafter \@secondoftwo
 \fi
}%
\providecommand \@ifx [1]{%
 \ifx #1\expandafter \@firstoftwo
 \else \expandafter \@secondoftwo
 \fi
}%
\providecommand \natexlab [1]{#1}%
\providecommand \enquote  [1]{``#1''}%
\providecommand \bibnamefont  [1]{#1}%
\providecommand \bibfnamefont [1]{#1}%
\providecommand \citenamefont [1]{#1}%
\providecommand \href@noop [0]{\@secondoftwo}%
\providecommand \href [0]{\begingroup \@sanitize@url \@href}%
\providecommand \@href[1]{\@@startlink{#1}\@@href}%
\providecommand \@@href[1]{\endgroup#1\@@endlink}%
\providecommand \@sanitize@url [0]{\catcode `\\12\catcode `\$12\catcode
  `\&12\catcode `\#12\catcode `\^12\catcode `\_12\catcode `\%12\relax}%
\providecommand \@@startlink[1]{}%
\providecommand \@@endlink[0]{}%
\providecommand \url  [0]{\begingroup\@sanitize@url \@url }%
\providecommand \@url [1]{\endgroup\@href {#1}{\urlprefix }}%
\providecommand \urlprefix  [0]{URL }%
\providecommand \Eprint [0]{\href }%
\providecommand \doibase [0]{https://doi.org/}%
\providecommand \selectlanguage [0]{\@gobble}%
\providecommand \bibinfo  [0]{\@secondoftwo}%
\providecommand \bibfield  [0]{\@secondoftwo}%
\providecommand \translation [1]{[#1]}%
\providecommand \BibitemOpen [0]{}%
\providecommand \bibitemStop [0]{}%
\providecommand \bibitemNoStop [0]{.\EOS\space}%
\providecommand \EOS [0]{\spacefactor3000\relax}%
\providecommand \BibitemShut  [1]{\csname bibitem#1\endcsname}%
\let\auto@bib@innerbib\@empty
\bibitem [{\citenamefont {Lenard}(1902)}]{Lenard1902}%
  \BibitemOpen
  \bibfield  {author} {\bibinfo {author} {\bibfnamefont {P.}~\bibnamefont
  {Lenard}},\ }\bibfield  {title} {\bibinfo {title} {{\"U}ber die
  lichtelektrische {W}irkung},\ }\href@noop {} {\bibfield  {journal} {\bibinfo
  {journal} {Annalen der Physik}\ }\textbf {\bibinfo {volume} {313}},\ \bibinfo
  {pages} {149} (\bibinfo {year} {1902})}\BibitemShut {NoStop}%
\bibitem [{\citenamefont {Einstein}(1905)}]{Einstein1905}%
  \BibitemOpen
  \bibfield  {author} {\bibinfo {author} {\bibfnamefont {A.}~\bibnamefont
  {Einstein}},\ }\bibfield  {title} {\bibinfo {title} {{\"U}ber einen die
  {E}rzeugung und {V}erwandlung des {L}ichtes betreffenden heuristischen
  {G}esichtspunkt},\ }\href@noop {} {\bibfield  {journal} {\bibinfo  {journal}
  {Annalen der Physik}\ }\textbf {\bibinfo {volume} {322}},\ \bibinfo {pages}
  {132} (\bibinfo {year} {1905})}\BibitemShut {NoStop}%
\bibitem [{\citenamefont {Perera}\ \emph {et~al.}(1992)\citenamefont {Perera},
  \citenamefont {Sherriff}, \citenamefont {Francombe},\ and\ \citenamefont
  {Devaty}}]{Perera1992}%
  \BibitemOpen
  \bibfield  {author} {\bibinfo {author} {\bibfnamefont {A.}~\bibnamefont
  {Perera}}, \bibinfo {author} {\bibfnamefont {R.}~\bibnamefont {Sherriff}},
  \bibinfo {author} {\bibfnamefont {M.}~\bibnamefont {Francombe}},\ and\
  \bibinfo {author} {\bibfnamefont {R.}~\bibnamefont {Devaty}},\ }\bibfield
  {title} {\bibinfo {title} {Far infrared photoelectric thresholds of extrinsic
  semiconductor photocathodes},\ }\href@noop {} {\bibfield  {journal} {\bibinfo
   {journal} {Appl. Phys. Lett.}\ }\textbf {\bibinfo {volume} {60}},\ \bibinfo
  {pages} {3168} (\bibinfo {year} {1992})}\BibitemShut {NoStop}%
\bibitem [{\citenamefont {Perera}\ \emph {et~al.}(1995)\citenamefont {Perera},
  \citenamefont {Yuan},\ and\ \citenamefont {Francombe}}]{Perera1995}%
  \BibitemOpen
  \bibfield  {author} {\bibinfo {author} {\bibfnamefont {A.}~\bibnamefont
  {Perera}}, \bibinfo {author} {\bibfnamefont {H.}~\bibnamefont {Yuan}},\ and\
  \bibinfo {author} {\bibfnamefont {M.}~\bibnamefont {Francombe}},\ }\bibfield
  {title} {\bibinfo {title} {Homojunction internal photoemission far-infrared
  detectors: photoresponse performance analysis},\ }\href@noop {} {\bibfield
  {journal} {\bibinfo  {journal} {J. Appl. Phys.}\ }\textbf {\bibinfo {volume}
  {77}},\ \bibinfo {pages} {915} (\bibinfo {year} {1995})}\BibitemShut
  {NoStop}%
\bibitem [{\citenamefont {Matsik}\ \emph {et~al.}(2003)\citenamefont {Matsik},
  \citenamefont {Rinzan}, \citenamefont {Perera}, \citenamefont {Liu},
  \citenamefont {Wasilewski},\ and\ \citenamefont {Buchanan}}]{Matsik2003}%
  \BibitemOpen
  \bibfield  {author} {\bibinfo {author} {\bibfnamefont {S.}~\bibnamefont
  {Matsik}}, \bibinfo {author} {\bibfnamefont {M.}~\bibnamefont {Rinzan}},
  \bibinfo {author} {\bibfnamefont {A.}~\bibnamefont {Perera}}, \bibinfo
  {author} {\bibfnamefont {H.}~\bibnamefont {Liu}}, \bibinfo {author}
  {\bibfnamefont {Z.}~\bibnamefont {Wasilewski}},\ and\ \bibinfo {author}
  {\bibfnamefont {M.}~\bibnamefont {Buchanan}},\ }\bibfield  {title} {\bibinfo
  {title} {Cutoff tailorability of heterojunction terahertz detectors},\
  }\href@noop {} {\bibfield  {journal} {\bibinfo  {journal} {Appl. Phys.
  Lett.}\ }\textbf {\bibinfo {volume} {82}},\ \bibinfo {pages} {139} (\bibinfo
  {year} {2003})}\BibitemShut {NoStop}%
\bibitem [{\citenamefont {Perera}\ \emph {et~al.}(2008)\citenamefont {Perera},
  \citenamefont {Ariyawansa}, \citenamefont {Jayaweera}, \citenamefont
  {Matsik}, \citenamefont {Buchanan},\ and\ \citenamefont {Liu}}]{Perera2008}%
  \BibitemOpen
  \bibfield  {author} {\bibinfo {author} {\bibfnamefont {A.}~\bibnamefont
  {Perera}}, \bibinfo {author} {\bibfnamefont {G.}~\bibnamefont {Ariyawansa}},
  \bibinfo {author} {\bibfnamefont {P.}~\bibnamefont {Jayaweera}}, \bibinfo
  {author} {\bibfnamefont {S.}~\bibnamefont {Matsik}}, \bibinfo {author}
  {\bibfnamefont {M.}~\bibnamefont {Buchanan}},\ and\ \bibinfo {author}
  {\bibfnamefont {H.}~\bibnamefont {Liu}},\ }\bibfield  {title} {\bibinfo
  {title} {Semiconductor terahertz detectors and absorption enhancement using
  plasmons},\ }\href@noop {} {\bibfield  {journal} {\bibinfo  {journal}
  {Microel. J.}\ }\textbf {\bibinfo {volume} {39}},\ \bibinfo {pages} {601}
  (\bibinfo {year} {2008})}\BibitemShut {NoStop}%
\bibitem [{\citenamefont {Shen}(2000)}]{Shen2000}%
  \BibitemOpen
  \bibfield  {author} {\bibinfo {author} {\bibfnamefont {W.}~\bibnamefont
  {Shen}},\ }\bibfield  {title} {\bibinfo {title} {Recent progress in mid-and
  far-infrared semiconductor detectors},\ }\href@noop {} {\bibfield  {journal}
  {\bibinfo  {journal} {Intern. J. of Infrared and Millimeter Waves}\ }\textbf
  {\bibinfo {volume} {21}},\ \bibinfo {pages} {1739} (\bibinfo {year}
  {2000})}\BibitemShut {NoStop}%
\bibitem [{\citenamefont {Lao}\ \emph {et~al.}(2014)\citenamefont {Lao},
  \citenamefont {Perera}, \citenamefont {Li}, \citenamefont {Khanna},
  \citenamefont {Linfield},\ and\ \citenamefont {Liu}}]{Lao2014}%
  \BibitemOpen
  \bibfield  {author} {\bibinfo {author} {\bibfnamefont {Y.-F.}\ \bibnamefont
  {Lao}}, \bibinfo {author} {\bibfnamefont {A.~U.}\ \bibnamefont {Perera}},
  \bibinfo {author} {\bibfnamefont {L.}~\bibnamefont {Li}}, \bibinfo {author}
  {\bibfnamefont {S.}~\bibnamefont {Khanna}}, \bibinfo {author} {\bibfnamefont
  {E.}~\bibnamefont {Linfield}},\ and\ \bibinfo {author} {\bibfnamefont
  {H.}~\bibnamefont {Liu}},\ }\bibfield  {title} {\bibinfo {title} {Tunable
  hot-carrier photodetection beyond the bandgap spectral limit},\ }\href@noop
  {} {\bibfield  {journal} {\bibinfo  {journal} {Nature Photonics}\ }\textbf
  {\bibinfo {volume} {8}},\ \bibinfo {pages} {412} (\bibinfo {year}
  {2014})}\BibitemShut {NoStop}%
\bibitem [{\citenamefont {Bai}\ \emph {et~al.}(2018)\citenamefont {Bai},
  \citenamefont {Zhang}, \citenamefont {Guo}, \citenamefont {Fu}, \citenamefont
  {Cao},\ and\ \citenamefont {Shen}}]{Bai2018}%
  \BibitemOpen
  \bibfield  {author} {\bibinfo {author} {\bibfnamefont {P.}~\bibnamefont
  {Bai}}, \bibinfo {author} {\bibfnamefont {Y.}~\bibnamefont {Zhang}}, \bibinfo
  {author} {\bibfnamefont {X.}~\bibnamefont {Guo}}, \bibinfo {author}
  {\bibfnamefont {Z.}~\bibnamefont {Fu}}, \bibinfo {author} {\bibfnamefont
  {J.}~\bibnamefont {Cao}},\ and\ \bibinfo {author} {\bibfnamefont
  {W.}~\bibnamefont {Shen}},\ }\bibfield  {title} {\bibinfo {title}
  {Realization of the high-performance {TH}z {G}a{A}s homojunction detector
  below the frequency of {R}eststrahlen band},\ }\href@noop {} {\bibfield
  {journal} {\bibinfo  {journal} {Appl. Phys. Lett.}\ }\textbf {\bibinfo
  {volume} {113}},\ \bibinfo {pages} {241102} (\bibinfo {year}
  {2018})}\BibitemShut {NoStop}%
\bibitem [{\citenamefont {Tamm}\ and\ \citenamefont
  {Schubin}(1931)}]{Tamm1931}%
  \BibitemOpen
  \bibfield  {author} {\bibinfo {author} {\bibfnamefont {I.}~\bibnamefont
  {Tamm}}\ and\ \bibinfo {author} {\bibfnamefont {S.}~\bibnamefont {Schubin}},\
  }\bibfield  {title} {\bibinfo {title} {Zur {T}heorie des {P}hotoeffektes an
  {M}etallen},\ }\href@noop {} {\bibfield  {journal} {\bibinfo  {journal}
  {Zeitschrift f\"ur Physik}\ }\textbf {\bibinfo {volume} {68}},\ \bibinfo
  {pages} {97} (\bibinfo {year} {1931})}\BibitemShut {NoStop}%
\bibitem [{\citenamefont {Mitchell}(1934)}]{Mitchell1934}%
  \BibitemOpen
  \bibfield  {author} {\bibinfo {author} {\bibfnamefont {K.}~\bibnamefont
  {Mitchell}},\ }\bibfield  {title} {\bibinfo {title} {The theory of the
  surface photoelectric effect in metals -- {I}},\ }\href@noop {} {\bibfield
  {journal} {\bibinfo  {journal} {Proc. of the Poyal Society A}\ }\textbf
  {\bibinfo {volume} {146}},\ \bibinfo {pages} {442} (\bibinfo {year}
  {1934})}\BibitemShut {NoStop}%
\bibitem [{\citenamefont {Levine}(1993)}]{Levine1993}%
  \BibitemOpen
  \bibfield  {author} {\bibinfo {author} {\bibfnamefont {B.~F.}\ \bibnamefont
  {Levine}},\ }\bibfield  {title} {\bibinfo {title} {Quantum-well infrared
  photodetectors},\ }\href@noop {} {\bibfield  {journal} {\bibinfo  {journal}
  {J. Appl. Phys.}\ }\textbf {\bibinfo {volume} {74}},\ \bibinfo {pages} {R1}
  (\bibinfo {year} {1993})}\BibitemShut {NoStop}%
\bibitem [{\citenamefont {Stiff-Roberts}(2009)}]{Stiff2009}%
  \BibitemOpen
  \bibfield  {author} {\bibinfo {author} {\bibfnamefont {A.~D.}\ \bibnamefont
  {Stiff-Roberts}},\ }\bibfield  {title} {\bibinfo {title} {Quantum-dot
  infrared photodetectors: a review},\ }\href@noop {} {\bibfield  {journal}
  {\bibinfo  {journal} {J. of Nanophotonics}\ }\textbf {\bibinfo {volume}
  {3}},\ \bibinfo {pages} {031607} (\bibinfo {year} {2009})}\BibitemShut
  {NoStop}%
\bibitem [{\citenamefont {Cates}\ \emph {et~al.}(1998)\citenamefont {Cates},
  \citenamefont {Briceno}, \citenamefont {Sherwin}, \citenamefont {Maranowski},
  \citenamefont {Campman},\ and\ \citenamefont {Gossard}}]{Cates1998}%
  \BibitemOpen
  \bibfield  {author} {\bibinfo {author} {\bibfnamefont {C.~L.}\ \bibnamefont
  {Cates}}, \bibinfo {author} {\bibfnamefont {G.}~\bibnamefont {Briceno}},
  \bibinfo {author} {\bibfnamefont {M.~S.}\ \bibnamefont {Sherwin}}, \bibinfo
  {author} {\bibfnamefont {K.~D.}\ \bibnamefont {Maranowski}}, \bibinfo
  {author} {\bibfnamefont {K.}~\bibnamefont {Campman}},\ and\ \bibinfo {author}
  {\bibfnamefont {A.~C.}\ \bibnamefont {Gossard}},\ }\bibfield  {title}
  {\bibinfo {title} {A concept for a tunable antenna-coupled intersubband
  terahertz ({TACIT}) detector},\ }\href@noop {} {\bibfield  {journal}
  {\bibinfo  {journal} {Physica E}\ }\textbf {\bibinfo {volume} {2}},\ \bibinfo
  {pages} {463} (\bibinfo {year} {1998})}\BibitemShut {NoStop}%
\bibitem [{\citenamefont {Michailow}\ \emph
  {et~al.}(2022{\natexlab{a}})\citenamefont {Michailow}, \citenamefont
  {Spencer}, \citenamefont {Almond}, \citenamefont {Kindness}, \citenamefont
  {Wallis}, \citenamefont {Mitchell}, \citenamefont {Degl'Innocenti},
  \citenamefont {Mikhailov}, \citenamefont {Beere},\ and\ \citenamefont
  {Ritchie}}]{Michailow22}%
  \BibitemOpen
  \bibfield  {author} {\bibinfo {author} {\bibfnamefont {W.}~\bibnamefont
  {Michailow}}, \bibinfo {author} {\bibfnamefont {P.}~\bibnamefont {Spencer}},
  \bibinfo {author} {\bibfnamefont {N.~W.}\ \bibnamefont {Almond}}, \bibinfo
  {author} {\bibfnamefont {S.~J.}\ \bibnamefont {Kindness}}, \bibinfo {author}
  {\bibfnamefont {R.}~\bibnamefont {Wallis}}, \bibinfo {author} {\bibfnamefont
  {T.~A.}\ \bibnamefont {Mitchell}}, \bibinfo {author} {\bibfnamefont
  {R.}~\bibnamefont {Degl'Innocenti}}, \bibinfo {author} {\bibfnamefont
  {S.~A.}\ \bibnamefont {Mikhailov}}, \bibinfo {author} {\bibfnamefont {H.~E.}\
  \bibnamefont {Beere}},\ and\ \bibinfo {author} {\bibfnamefont {D.~A.}\
  \bibnamefont {Ritchie}},\ }\bibfield  {title} {\bibinfo {title} {An in-plane
  photoelectric effect in two-dimensional electron systems for terahertz
  detection},\ }\href@noop {} {\bibfield  {journal} {\bibinfo  {journal} {Sci.
  Adv.}\ }\textbf {\bibinfo {volume} {8}},\ \bibinfo {pages} {eabi8398}
  (\bibinfo {year} {2022}{\natexlab{a}})}\BibitemShut {NoStop}%
\bibitem [{\citenamefont {Degl'Innocenti}\ \emph {et~al.}(2017)\citenamefont
  {Degl'Innocenti}, \citenamefont {Xiao}, \citenamefont {Kindness},
  \citenamefont {Kamboj}, \citenamefont {Wei}, \citenamefont
  {Braeuninger-Weimer}, \citenamefont {Nakanishi}, \citenamefont {Aria},
  \citenamefont {Hofmann}, \citenamefont {Beere},\ and\ \citenamefont
  {Ritchie}}]{DeglInnocenti17}%
  \BibitemOpen
  \bibfield  {author} {\bibinfo {author} {\bibfnamefont {R.}~\bibnamefont
  {Degl'Innocenti}}, \bibinfo {author} {\bibfnamefont {L.}~\bibnamefont
  {Xiao}}, \bibinfo {author} {\bibfnamefont {S.~J.}\ \bibnamefont {Kindness}},
  \bibinfo {author} {\bibfnamefont {V.~S.}\ \bibnamefont {Kamboj}}, \bibinfo
  {author} {\bibfnamefont {B.}~\bibnamefont {Wei}}, \bibinfo {author}
  {\bibfnamefont {P.}~\bibnamefont {Braeuninger-Weimer}}, \bibinfo {author}
  {\bibfnamefont {K.}~\bibnamefont {Nakanishi}}, \bibinfo {author}
  {\bibfnamefont {A.~I.}\ \bibnamefont {Aria}}, \bibinfo {author}
  {\bibfnamefont {S.}~\bibnamefont {Hofmann}}, \bibinfo {author} {\bibfnamefont
  {H.~E.}\ \bibnamefont {Beere}},\ and\ \bibinfo {author} {\bibfnamefont
  {D.~A.}\ \bibnamefont {Ritchie}},\ }\bibfield  {title} {\bibinfo {title}
  {Bolometric detection of terahertz quantum cascade laser radiation with
  graphene-plasmonic antenna arrays},\ }\href@noop {} {\bibfield  {journal}
  {\bibinfo  {journal} {Journal of Physics D: Applied Physics}\ }\textbf
  {\bibinfo {volume} {50}},\ \bibinfo {pages} {174001} (\bibinfo {year}
  {2017})}\BibitemShut {NoStop}%
\bibitem [{\citenamefont {Gabor}\ \emph {et~al.}(2011)\citenamefont {Gabor},
  \citenamefont {Song}, \citenamefont {Ma}, \citenamefont {Nair}, \citenamefont
  {Taychatanapat}, \citenamefont {Watanabe}, \citenamefont {Taniguchi},
  \citenamefont {Levitov},\ and\ \citenamefont {Jarillo-Herrero}}]{Gabor11}%
  \BibitemOpen
  \bibfield  {author} {\bibinfo {author} {\bibfnamefont {N.~M.}\ \bibnamefont
  {Gabor}}, \bibinfo {author} {\bibfnamefont {J.~C.}\ \bibnamefont {Song}},
  \bibinfo {author} {\bibfnamefont {Q.}~\bibnamefont {Ma}}, \bibinfo {author}
  {\bibfnamefont {N.~L.}\ \bibnamefont {Nair}}, \bibinfo {author}
  {\bibfnamefont {T.}~\bibnamefont {Taychatanapat}}, \bibinfo {author}
  {\bibfnamefont {K.}~\bibnamefont {Watanabe}}, \bibinfo {author}
  {\bibfnamefont {T.}~\bibnamefont {Taniguchi}}, \bibinfo {author}
  {\bibfnamefont {L.~S.}\ \bibnamefont {Levitov}},\ and\ \bibinfo {author}
  {\bibfnamefont {P.}~\bibnamefont {Jarillo-Herrero}},\ }\bibfield  {title}
  {\bibinfo {title} {Hot carrier-assisted intrinsic photoresponse in
  graphene},\ }\href@noop {} {\bibfield  {journal} {\bibinfo  {journal}
  {Science}\ }\textbf {\bibinfo {volume} {334}},\ \bibinfo {pages} {648}
  (\bibinfo {year} {2011})}\BibitemShut {NoStop}%
\bibitem [{\citenamefont {Viti}\ \emph {et~al.}(2016)\citenamefont {Viti},
  \citenamefont {Hu}, \citenamefont {Coquillat}, \citenamefont {Politano},
  \citenamefont {Consejo}, \citenamefont {W.},\ and\ \citenamefont
  {Vitiello}}]{Viti16}%
  \BibitemOpen
  \bibfield  {author} {\bibinfo {author} {\bibfnamefont {L.}~\bibnamefont
  {Viti}}, \bibinfo {author} {\bibfnamefont {J.}~\bibnamefont {Hu}}, \bibinfo
  {author} {\bibfnamefont {D.}~\bibnamefont {Coquillat}}, \bibinfo {author}
  {\bibfnamefont {A.}~\bibnamefont {Politano}}, \bibinfo {author}
  {\bibfnamefont {C.}~\bibnamefont {Consejo}}, \bibinfo {author} {\bibfnamefont
  {K.}~\bibnamefont {W.}},\ and\ \bibinfo {author} {\bibfnamefont {M.~S.}\
  \bibnamefont {Vitiello}},\ }\bibfield  {title} {\bibinfo {title}
  {Heterostructured h{BN}-{BP}-h{BN} nanodetectors at terahertz frequencies},\
  }\href@noop {} {\bibfield  {journal} {\bibinfo  {journal} {Advanced
  Materials}\ }\textbf {\bibinfo {volume} {28}},\ \bibinfo {pages} {7390 }
  (\bibinfo {year} {2016})}\BibitemShut {NoStop}%
\bibitem [{\citenamefont {Castilla}\ \emph {et~al.}(2019)\citenamefont
  {Castilla}, \citenamefont {Terres}, \citenamefont {Autore}, \citenamefont
  {Viti}, \citenamefont {Li}, \citenamefont {Nikitin}, \citenamefont
  {Vangelidis}, \citenamefont {Watanabe}, \citenamefont {Taniguchi},
  \citenamefont {Lidorikis}, \citenamefont {Vitiello}, \citenamefont
  {Hillenbrand}, \citenamefont {Tielrooij},\ and\ \citenamefont
  {Koppens}}]{Castilla19}%
  \BibitemOpen
  \bibfield  {author} {\bibinfo {author} {\bibfnamefont {S.}~\bibnamefont
  {Castilla}}, \bibinfo {author} {\bibfnamefont {B.}~\bibnamefont {Terres}},
  \bibinfo {author} {\bibfnamefont {M.}~\bibnamefont {Autore}}, \bibinfo
  {author} {\bibfnamefont {L.}~\bibnamefont {Viti}}, \bibinfo {author}
  {\bibfnamefont {J.}~\bibnamefont {Li}}, \bibinfo {author} {\bibfnamefont
  {A.~Y.}\ \bibnamefont {Nikitin}}, \bibinfo {author} {\bibfnamefont
  {I.}~\bibnamefont {Vangelidis}}, \bibinfo {author} {\bibfnamefont
  {K.}~\bibnamefont {Watanabe}}, \bibinfo {author} {\bibfnamefont
  {T.}~\bibnamefont {Taniguchi}}, \bibinfo {author} {\bibfnamefont
  {E.}~\bibnamefont {Lidorikis}}, \bibinfo {author} {\bibfnamefont {M.~S.}\
  \bibnamefont {Vitiello}}, \bibinfo {author} {\bibfnamefont {R.}~\bibnamefont
  {Hillenbrand}}, \bibinfo {author} {\bibfnamefont {K.-J.}\ \bibnamefont
  {Tielrooij}},\ and\ \bibinfo {author} {\bibfnamefont {F.~H.~L.}\ \bibnamefont
  {Koppens}},\ }\bibfield  {title} {\bibinfo {title} {Fast and sensitive
  terahertz detection using an antenna-integrated graphene pn junction},\
  }\href@noop {} {\bibfield  {journal} {\bibinfo  {journal} {Nano Letters}\
  }\textbf {\bibinfo {volume} {19}},\ \bibinfo {pages} {2765 } (\bibinfo {year}
  {2019})}\BibitemShut {NoStop}%
\bibitem [{\citenamefont {Viti}\ \emph {et~al.}(2021)\citenamefont {Viti},
  \citenamefont {Cadore}, \citenamefont {Yang}, \citenamefont {Vorobiev},
  \citenamefont {Muench}, \citenamefont {Watanabe}, \citenamefont {Taniguchi},
  \citenamefont {Stake}, \citenamefont {Ferrari},\ and\ \citenamefont
  {Vitiello}}]{Viti20}%
  \BibitemOpen
  \bibfield  {author} {\bibinfo {author} {\bibfnamefont {L.}~\bibnamefont
  {Viti}}, \bibinfo {author} {\bibfnamefont {A.~R.}\ \bibnamefont {Cadore}},
  \bibinfo {author} {\bibfnamefont {X.}~\bibnamefont {Yang}}, \bibinfo {author}
  {\bibfnamefont {A.}~\bibnamefont {Vorobiev}}, \bibinfo {author}
  {\bibfnamefont {J.~E.}\ \bibnamefont {Muench}}, \bibinfo {author}
  {\bibfnamefont {K.}~\bibnamefont {Watanabe}}, \bibinfo {author}
  {\bibfnamefont {T.}~\bibnamefont {Taniguchi}}, \bibinfo {author}
  {\bibfnamefont {J.}~\bibnamefont {Stake}}, \bibinfo {author} {\bibfnamefont
  {A.~C.}\ \bibnamefont {Ferrari}},\ and\ \bibinfo {author} {\bibfnamefont
  {M.~S.}\ \bibnamefont {Vitiello}},\ }\bibfield  {title} {\bibinfo {title}
  {Thermoelectric graphene photodetectors with sub-nanosecond response times at
  terahertz frequencies},\ }\href@noop {} {\bibfield  {journal} {\bibinfo
  {journal} {Nanophotonics}\ }\textbf {\bibinfo {volume} {10}},\ \bibinfo
  {pages} {89} (\bibinfo {year} {2021})}\BibitemShut {NoStop}%
\bibitem [{\citenamefont {Dyakonov}\ and\ \citenamefont
  {Shur}(1996)}]{Dyakonov96}%
  \BibitemOpen
  \bibfield  {author} {\bibinfo {author} {\bibfnamefont {M.~I.}\ \bibnamefont
  {Dyakonov}}\ and\ \bibinfo {author} {\bibfnamefont {M.}~\bibnamefont
  {Shur}},\ }\bibfield  {title} {\bibinfo {title} {Detection, mixing, and
  frequency multiplication of terahertz radiation by two dimensional electronic
  fluid},\ }\href@noop {} {\bibfield  {journal} {\bibinfo  {journal} {IEEE
  Trans. Electron. Dev.}\ }\textbf {\bibinfo {volume} {43}},\ \bibinfo {pages}
  {380} (\bibinfo {year} {1996})}\BibitemShut {NoStop}%
\bibitem [{\citenamefont {Viti}\ \emph {et~al.}(2015)\citenamefont {Viti},
  \citenamefont {Hu}, \citenamefont {Coquillat}, \citenamefont {W.},
  \citenamefont {Tredicucci}, \citenamefont {Politano},\ and\ \citenamefont
  {Vitiello}}]{Viti15}%
  \BibitemOpen
  \bibfield  {author} {\bibinfo {author} {\bibfnamefont {L.}~\bibnamefont
  {Viti}}, \bibinfo {author} {\bibfnamefont {J.}~\bibnamefont {Hu}}, \bibinfo
  {author} {\bibfnamefont {D.}~\bibnamefont {Coquillat}}, \bibinfo {author}
  {\bibfnamefont {K.}~\bibnamefont {W.}}, \bibinfo {author} {\bibfnamefont
  {A.}~\bibnamefont {Tredicucci}}, \bibinfo {author} {\bibfnamefont
  {A.}~\bibnamefont {Politano}},\ and\ \bibinfo {author} {\bibfnamefont
  {M.~S.}\ \bibnamefont {Vitiello}},\ }\bibfield  {title} {\bibinfo {title}
  {Black phosphorus terahertz photodetectors},\ }\href@noop {} {\bibfield
  {journal} {\bibinfo  {journal} {Advanced Materials}\ }\textbf {\bibinfo
  {volume} {27}},\ \bibinfo {pages} {5567 } (\bibinfo {year}
  {2015})}\BibitemShut {NoStop}%
\bibitem [{\citenamefont {Bandurin}\ \emph {et~al.}(2018)\citenamefont
  {Bandurin}, \citenamefont {Svintsov}, \citenamefont {Gayduchenko},
  \citenamefont {Xu}, \citenamefont {Principi}, \citenamefont {Moskotin},
  \citenamefont {Tretyakov}, \citenamefont {Yagodkin}, \citenamefont {Zhukov},
  \citenamefont {Taniguchi}, \citenamefont {Watanabe}, \citenamefont
  {Grigorieva}, \citenamefont {Polini}, \citenamefont {Goltsman}, \citenamefont
  {Geim},\ and\ \citenamefont {Fedorov}}]{Bandurin18}%
  \BibitemOpen
  \bibfield  {author} {\bibinfo {author} {\bibfnamefont {D.~A.}\ \bibnamefont
  {Bandurin}}, \bibinfo {author} {\bibfnamefont {D.}~\bibnamefont {Svintsov}},
  \bibinfo {author} {\bibfnamefont {I.}~\bibnamefont {Gayduchenko}}, \bibinfo
  {author} {\bibfnamefont {S.~G.}\ \bibnamefont {Xu}}, \bibinfo {author}
  {\bibfnamefont {A.}~\bibnamefont {Principi}}, \bibinfo {author}
  {\bibfnamefont {M.}~\bibnamefont {Moskotin}}, \bibinfo {author}
  {\bibfnamefont {I.}~\bibnamefont {Tretyakov}}, \bibinfo {author}
  {\bibfnamefont {D.}~\bibnamefont {Yagodkin}}, \bibinfo {author}
  {\bibfnamefont {S.}~\bibnamefont {Zhukov}}, \bibinfo {author} {\bibfnamefont
  {T.}~\bibnamefont {Taniguchi}}, \bibinfo {author} {\bibfnamefont
  {K.}~\bibnamefont {Watanabe}}, \bibinfo {author} {\bibfnamefont {I.~V.}\
  \bibnamefont {Grigorieva}}, \bibinfo {author} {\bibfnamefont
  {M.}~\bibnamefont {Polini}}, \bibinfo {author} {\bibfnamefont {G.~N.}\
  \bibnamefont {Goltsman}}, \bibinfo {author} {\bibfnamefont {A.~K.}\
  \bibnamefont {Geim}},\ and\ \bibinfo {author} {\bibfnamefont
  {G.}~\bibnamefont {Fedorov}},\ }\bibfield  {title} {\bibinfo {title}
  {Resonant terahertz detection using graphene plasmons},\ }\href@noop {}
  {\bibfield  {journal} {\bibinfo  {journal} {Nature Communications}\ }\textbf
  {\bibinfo {volume} {9}},\ \bibinfo {pages} {5392} (\bibinfo {year}
  {2018})}\BibitemShut {NoStop}%
\bibitem [{\citenamefont {Sun}\ \emph {et~al.}(2012)\citenamefont {Sun},
  \citenamefont {Sun}, \citenamefont {Wu}, \citenamefont {Cai}, \citenamefont
  {Qin},\ and\ \citenamefont {Zhang}}]{Sun12a}%
  \BibitemOpen
  \bibfield  {author} {\bibinfo {author} {\bibfnamefont {J.~D.}\ \bibnamefont
  {Sun}}, \bibinfo {author} {\bibfnamefont {Y.~F.}\ \bibnamefont {Sun}},
  \bibinfo {author} {\bibfnamefont {D.~M.}\ \bibnamefont {Wu}}, \bibinfo
  {author} {\bibfnamefont {Y.}~\bibnamefont {Cai}}, \bibinfo {author}
  {\bibfnamefont {H.}~\bibnamefont {Qin}},\ and\ \bibinfo {author}
  {\bibfnamefont {B.~S.}\ \bibnamefont {Zhang}},\ }\bibfield  {title} {\bibinfo
  {title} {High-responsivity, low-noise, room-temperature, self-mixing
  terahertz detector realized using floating antennas on a {G}a{N}-based
  field-effect transistor},\ }\href@noop {} {\bibfield  {journal} {\bibinfo
  {journal} {Appl. Phys. Lett.}\ }\textbf {\bibinfo {volume} {100}},\ \bibinfo
  {pages} {013506} (\bibinfo {year} {2012})}\BibitemShut {NoStop}%
\bibitem [{\citenamefont {Stern}(1967)}]{Stern67}%
  \BibitemOpen
  \bibfield  {author} {\bibinfo {author} {\bibfnamefont {F.}~\bibnamefont
  {Stern}},\ }\bibfield  {title} {\bibinfo {title} {Polarizability of a
  two-dimensional electron gas},\ }\href@noop {} {\bibfield  {journal}
  {\bibinfo  {journal} {Phys. Rev. Lett.}\ }\textbf {\bibinfo {volume} {18}},\
  \bibinfo {pages} {546} (\bibinfo {year} {1967})}\BibitemShut {NoStop}%
\bibitem [{\citenamefont {Chaplik}(1972)}]{Chaplik72}%
  \BibitemOpen
  \bibfield  {author} {\bibinfo {author} {\bibfnamefont {A.~V.}\ \bibnamefont
  {Chaplik}},\ }\bibfield  {title} {\bibinfo {title} {Possible crystallization
  of charge carriers in low-density inversion layers},\ }\href@noop {}
  {\bibfield  {journal} {\bibinfo  {journal} {Zh. Eksp. Teor. Fiz.}\ }\textbf
  {\bibinfo {volume} {62}},\ \bibinfo {pages} {746} (\bibinfo {year} {1972})},\
  \bibinfo {note} {[Sov. Phys.--JETP {\bf 35}, 395-398 (1972)]}\BibitemShut
  {NoStop}%
\bibitem [{\citenamefont {Landau}\ and\ \citenamefont
  {Lifshitz}(1994)}]{Landau3}%
  \BibitemOpen
  \bibfield  {author} {\bibinfo {author} {\bibfnamefont {L.~D.}\ \bibnamefont
  {Landau}}\ and\ \bibinfo {author} {\bibfnamefont {E.~M.}\ \bibnamefont
  {Lifshitz}},\ }\href@noop {} {\emph {\bibinfo {title} {Quantum mechanics
  (Non-relativistic theory)}}}\ (\bibinfo  {publisher} {Elsevier},\ \bibinfo
  {address} {Oxford},\ \bibinfo {year} {1994})\BibitemShut {NoStop}%
\bibitem [{\citenamefont {van Wees}\ \emph {et~al.}(1988)\citenamefont {van
  Wees}, \citenamefont {van Houten}, \citenamefont {Beenakker}, \citenamefont
  {Williamson}, \citenamefont {Kouwenhoven}, \citenamefont {van~der Marel},\
  and\ \citenamefont {Foxon}}]{Wees88}%
  \BibitemOpen
  \bibfield  {author} {\bibinfo {author} {\bibfnamefont {B.~J.}\ \bibnamefont
  {van Wees}}, \bibinfo {author} {\bibfnamefont {H.}~\bibnamefont {van
  Houten}}, \bibinfo {author} {\bibfnamefont {C.~W.~J.}\ \bibnamefont
  {Beenakker}}, \bibinfo {author} {\bibfnamefont {J.~G.}\ \bibnamefont
  {Williamson}}, \bibinfo {author} {\bibfnamefont {L.~P.}\ \bibnamefont
  {Kouwenhoven}}, \bibinfo {author} {\bibfnamefont {D.}~\bibnamefont {van~der
  Marel}},\ and\ \bibinfo {author} {\bibfnamefont {C.~T.}\ \bibnamefont
  {Foxon}},\ }\bibfield  {title} {\bibinfo {title} {Quantized conductance of
  point contacts in a two-dimensional electron gas},\ }\href@noop {} {\bibfield
   {journal} {\bibinfo  {journal} {Phys. Rev. Lett.}\ }\textbf {\bibinfo
  {volume} {60}},\ \bibinfo {pages} {848} (\bibinfo {year} {1988})}\BibitemShut
  {NoStop}%
\bibitem [{\citenamefont {Wharam}\ \emph {et~al.}(1988)\citenamefont {Wharam},
  \citenamefont {Thornton}, \citenamefont {Newbury}, \citenamefont {Pepper},
  \citenamefont {Ahmed}, \citenamefont {Frost}, \citenamefont {Hasko},
  \citenamefont {Peacock}, \citenamefont {Ritchie},\ and\ \citenamefont
  {Jones}}]{Wharam88}%
  \BibitemOpen
  \bibfield  {author} {\bibinfo {author} {\bibfnamefont {D.}~\bibnamefont
  {Wharam}}, \bibinfo {author} {\bibfnamefont {T.~J.}\ \bibnamefont
  {Thornton}}, \bibinfo {author} {\bibfnamefont {R.}~\bibnamefont {Newbury}},
  \bibinfo {author} {\bibfnamefont {M.}~\bibnamefont {Pepper}}, \bibinfo
  {author} {\bibfnamefont {H.}~\bibnamefont {Ahmed}}, \bibinfo {author}
  {\bibfnamefont {J.~E.~F.}\ \bibnamefont {Frost}}, \bibinfo {author}
  {\bibfnamefont {D.~G.}\ \bibnamefont {Hasko}}, \bibinfo {author}
  {\bibfnamefont {D.~C.}\ \bibnamefont {Peacock}}, \bibinfo {author}
  {\bibfnamefont {D.~A.}\ \bibnamefont {Ritchie}},\ and\ \bibinfo {author}
  {\bibfnamefont {G.~A.~C.}\ \bibnamefont {Jones}},\ }\bibfield  {title}
  {\bibinfo {title} {One-dimensional transport and the quantisation of the
  ballistic resistance},\ }\href@noop {} {\bibfield  {journal} {\bibinfo
  {journal} {J. Phys. C: Solid State Phys.}\ }\textbf {\bibinfo {volume}
  {21}},\ \bibinfo {pages} {L209} (\bibinfo {year} {1988})}\BibitemShut
  {NoStop}%
\bibitem [{\citenamefont {Michailow}\ \emph
  {et~al.}(2022{\natexlab{b}})\citenamefont {Michailow}, \citenamefont
  {Spencer}, \citenamefont {Almond}, \citenamefont {Kindness}, \citenamefont
  {Wallis}, \citenamefont {Mitchell}, \citenamefont {Degl'Innocenti},
  \citenamefont {Mikhailov}, \citenamefont {Beere},\ and\ \citenamefont
  {Ritchie}}]{Michailow22-data}%
  \BibitemOpen
  \bibfield  {author} {\bibinfo {author} {\bibfnamefont {W.}~\bibnamefont
  {Michailow}}, \bibinfo {author} {\bibfnamefont {P.}~\bibnamefont {Spencer}},
  \bibinfo {author} {\bibfnamefont {N.~W.}\ \bibnamefont {Almond}}, \bibinfo
  {author} {\bibfnamefont {S.~J.}\ \bibnamefont {Kindness}}, \bibinfo {author}
  {\bibfnamefont {R.}~\bibnamefont {Wallis}}, \bibinfo {author} {\bibfnamefont
  {T.~A.}\ \bibnamefont {Mitchell}}, \bibinfo {author} {\bibfnamefont
  {R.}~\bibnamefont {Degl'Innocenti}}, \bibinfo {author} {\bibfnamefont
  {S.~A.}\ \bibnamefont {Mikhailov}}, \bibinfo {author} {\bibfnamefont {H.~E.}\
  \bibnamefont {Beere}},\ and\ \bibinfo {author} {\bibfnamefont {D.~A.}\
  \bibnamefont {Ritchie}},\ }\bibfield  {title} {\bibinfo {title} {Research
  data supporting "{A}n in-plane photoelectric effect in two-dimensional
  electron systems for terahertz detection"},\ }\bibfield  {journal} {\bibinfo
  {journal} {Apollo - University of Cambridge Repository,}\ }\href
  {https://doi.org/10.17863/CAM.58046} {10.17863/CAM.58046} (\bibinfo {year}
  {2022}{\natexlab{b}})\BibitemShut {NoStop}%
\bibitem [{\citenamefont {Levin}\ \emph {et~al.}(2015)\citenamefont {Levin},
  \citenamefont {Gusev}, \citenamefont {Kvon}, \citenamefont {Bakarov},
  \citenamefont {Savostianova}, \citenamefont {Mikhailov}, \citenamefont
  {Rodyakina},\ and\ \citenamefont {Latyshev}}]{Levin15}%
  \BibitemOpen
  \bibfield  {author} {\bibinfo {author} {\bibfnamefont {A.~D.}\ \bibnamefont
  {Levin}}, \bibinfo {author} {\bibfnamefont {G.~M.}\ \bibnamefont {Gusev}},
  \bibinfo {author} {\bibfnamefont {Z.~D.}\ \bibnamefont {Kvon}}, \bibinfo
  {author} {\bibfnamefont {A.~K.}\ \bibnamefont {Bakarov}}, \bibinfo {author}
  {\bibfnamefont {N.~A.}\ \bibnamefont {Savostianova}}, \bibinfo {author}
  {\bibfnamefont {S.~A.}\ \bibnamefont {Mikhailov}}, \bibinfo {author}
  {\bibfnamefont {E.~E.}\ \bibnamefont {Rodyakina}},\ and\ \bibinfo {author}
  {\bibfnamefont {A.~V.}\ \bibnamefont {Latyshev}},\ }\bibfield  {title}
  {\bibinfo {title} {Giant microwave photo-conductance of a tunnel point
  contact with a bridged gate},\ }\href@noop {} {\bibfield  {journal} {\bibinfo
   {journal} {Appl. Phys. Lett.}\ }\textbf {\bibinfo {volume} {107}},\ \bibinfo
  {pages} {072112} (\bibinfo {year} {2015})}\BibitemShut {NoStop}%
\bibitem [{\citenamefont {McColl}\ \emph {et~al.}(1977)\citenamefont {McColl},
  \citenamefont {Hogges},\ and\ \citenamefont {Garber}}]{McColl77}%
  \BibitemOpen
  \bibfield  {author} {\bibinfo {author} {\bibfnamefont {M.}~\bibnamefont
  {McColl}}, \bibinfo {author} {\bibfnamefont {D.~T.}\ \bibnamefont {Hogges}},\
  and\ \bibinfo {author} {\bibfnamefont {W.~A.}\ \bibnamefont {Garber}},\
  }\bibfield  {title} {\bibinfo {title} {Submillimeter-wave detection with
  submicron-size {S}chottky-barrier diodes},\ }\href@noop {} {\bibfield
  {journal} {\bibinfo  {journal} {IEEE Trans. Microwave Theory and Techniques}\
  }\textbf {\bibinfo {volume} {MTT-25}},\ \bibinfo {pages} {463} (\bibinfo
  {year} {1977})}\BibitemShut {NoStop}%
\end{thebibliography}
\end{document}